\documentclass[12pt,prepring]{aastex}
\newcommand*{\teff}{$T_{\rm{eff}}$}
\newcommand*{\logg}{$\log~g$}
\newcommand*{\feh}{[Fe/H]}
\newcommand*{\gr}{$(g-r)_{0}$}

\usepackage{graphics,graphicx,amsmath,amssymb,amsthm,epsfig,lscape}
\shorttitle{SDSS Globular Cluster CN and CH Variations}
\shortauthors{Smolinski et al.}

\begin{document}

\title{A Survey of CN and CH Variations in Galactic Globular Clusters from SDSS Spectroscopy}

\bigskip
\author{Jason P. Smolinski}
\affil{Department of Physics \& Astronomy and JINA: Joint Institute for Nuclear
Astrophysics, Michigan State University, East Lansing, MI 48824, USA}
\email{smolin19@msu.edu}

\author{Sarah L. Martell} 
\affil{Astronomisches Rechen-Institut, Zentrum f$\rm{\ddot{u}}$r Astronomie der Universit$\rm{\ddot{a}}$t
Heidelberg, 69120 Heidelberg, Germany}
\email{martell@ari.uni-heidelberg.de}

\author{Timothy C. Beers, Young Sun Lee}
\affil{Department of Physics \& Astronomy and JINA: Joint Institute for Nuclear
Astrophysics, Michigan State University, East Lansing, MI 48824, USA}
\email{beers@pa.msu.edu, lee@pa.msu.edu} 

%\and 
%\author{}

\bigskip
\begin{abstract} 

We present a homogeneous survey of the CN and CH bandstrengths in eight
Galactic globular clusters observed during the course of the Sloan
Extension for Galactic Understanding and Exploration (SEGUE) sub-survey of
the SDSS. We confirm the existence of a bimodal CN distribution among RGB
stars in all of the clusters with metallicity greater than \feh ~$=-1.7$;
the lowest metallicity cluster with an observed CN bimodality is M53, with
\feh ~$\simeq -2.1$. There is also some evidence for individual CN groups
on the subgiant branches of M92, M2, and M13, and on the red giant branches
of M92 and NGC~5053. Finally, we quantify the correlation between overall
cluster metallicity and the slope of the CN bandstrength-luminosity plot as
a means of further demonstrating the level of CN-enrichment in cluster
giants. Our results agree well with previous studies reported in the
literature.

\end{abstract}

\keywords{Globular clusters: individual (M2, M3, M13, M15, M53, M71, M92,
NGC~5053) - Stars: abundances - Stars: evolution}

\section{Introduction}\label{secintro}

Standard models for the formation of Galactic globular clusters (GCs) have long
held that they should display little star-to-star variations in their
observed atmospheric elemental abundances. However, observations over the
past forty years \citep[beginning with][]{osb71} have repeatedly
shown this expectation is not fully met, with variations in carbon and
nitrogen abundance being commonly studied through the strengths of the
3883\,{\AA} CN and 4320\,{\AA} CH absorption bands. For clusters
of moderate to high metallicity (\feh ~$> -2.0$), significant scatter in
light-element abundances has been observed on the red giant branch
\citep[RGB; e.g.][]{nor79,sun81}, and in some cases even down to the main
sequence \citep[MS; e.g.][]{can98,har03a,bri04}.

The observed variations in carbon and nitrogen abundance are part of a
larger light-element pattern that involves enrichment in N, Na, and Mg
along with depletion in C, O, and Al, and is often studied in
correlated or anticorrelated abundance pairs (C-N, O-N, Mg-Al, Na-O,
etc.). \citet{gra04} reviews a number of these studies, and
\citet{car09b} dramatically increased the number of cluster stars
surveyed for these variations. There are two independent modes of
variation in globular cluster light-element abundances: a steady
decline in [C/Fe] and increase in [N/Fe] as stars evolve along the
RGB, and star-to-star variations in the light-element abundances at a
fixed luminosity, at all evolutionary phases.

Several hypotheses have been proposed to explain these observed
anomalies. The progressive abundance changes on the RGB are believed
to be the result of deep mixing within individual
stars \citep{swe79, cha94,den03}, beginning at the ``bump'' in the RGB
luminosity function  \citep{fus90, she03}. The hydrogen-burning shell proceeds
outward as a star evolves along the RGB, eventually \citep[for all
stars with M $< 2.5\rm{M}_{\odot}$;][]{gil89}
encountering the molecular-weight discontinuity left behind by the inward
reach of the convective envelope during first dredge-up \citep{tho67,
ibe68}. When this occurs, the shell's progress is delayed as its fusion
rate adjusts to the new chemical abundances, causing a loop in the star's
evolution along the RGB. In a population of coeval stars, this produces an
enhancement in the differential luminosity function. Once the shell begins
to proceed outward again, the molecular-weight gradient it experiences is
lower, and the process of deep mixing \citep{den03} begins to
operate, transporting material between the hydrogen-burning shell and
the surface and continuously adjusting surface carbon and nitrogen abundances.

Early studies of star-to-star light-element abundance variations
\citep[e.g.][]{sun81,lan85} suggested that deep mixing might be
responsible for the C-N variations at fixed luminosity as well as the
progressive abundance changes along the RGB. However, variations in
sodium, magnesium and aluminum are difficult to explain as a result of
mixing within RGB stars since they require higher temperatures than
are reached in the hydrogen-burning shell. Internal mixing is also not
a good explanation for light-element abundance variations in
main-sequence stars, since they do not have the ability to conduct
either high-temperature fusion or mass transport between their cores
and surfaces.

The presence of these abundance variations at all evolutionary phases implies
some form of unexpected enrichment within individual clusters prior
to, or shortly after, their formation. Some researchers have suggested
that the primordial gas cloud from which a given cluster formed may
have initially been chemically inhomogeneous
\citep[e.g.][]{coh78,pet80}. A variation on this hypothesis appeals to
cluster ``self-enrichment.'' Rather than assuming all stars in a given
cluster are co-eval, it is assumed that an additional population(s) of
stars formed, with compositions affected by the gas expelled by
supernovae and/or strong stellar winds from intermediate-mass
asymptotic giant branch (AGB) stars. While each hypothesis has its
merits and weaknesses \citep[see][for a complete review]{gra04}, no
single model accounts for the full set of observed light-element
abundance variations in GCs -- it remains possible that each may play
a role \citep[e.g.][]{mar08c,dec09}.

One of the strongest constraints on the proposed scenarios is the fact
that, in the majority of moderate- to high-metallicity GCs, the CN
abundance distribution is bimodal. The self-enrichment scenario
accommodates this observation most naturally, with the CN-weak and
CN-strong groups representing the first and second populations of stars to
have formed in the cluster, respectively. This idea has received increasing
recent support, as improved photometric measurements have revealed the
presence of multiple subgiant branches (SGBs) and main sequences in many
GCs \citep{bed04,pio07}. Spectroscopic abundance analyses have also begun
to reveal distinct sequences on the RGBs of some clusters \citep{mari08,
lar11}. In addition, studies of the spatial distributions of stars in
clusters with accurate photometry have revealed the presence of correlated
differences in $U-B$ colors \citep{car10a, kra11}, which suggests variations
in chemical compositions within the cluster.

In this paper we use available photometric and spectroscopic data from Data
Release 7 \citep[DR7;][]{aba09} of the Sloan Digital Sky Survey
\citep[SDSS;][]{yor00} to examine the CN and CH bandstrength
distributions for stars in eight GCs, including stars from the upper RGB
to, in some cases, 1--2 magnitudes below the main sequence turnoff (MSTO).
We show that there exists a clear bimodal distribution in CN bandstrengths
for clusters with \feh ~$\geq -2.0$. Other interesting CN bandstrength
variations are suggested to exist among the three clusters in our sample
with \feh ~$< -2.0$.

This paper is organized as follows. In Section \ref{secdata} we briefly
describe our data and the cluster membership selection process. In Section
\ref{seccn} we define the adopted CN and CH indices for stars in various stages of
evolution. We then compare the derived CN and CH distributions in Section
\ref{seccnch}. In Section \ref{secclustcorr} we search for any correlations
between these distributions with the global cluster parameters. Finally,
our results and implications are discussed in Section \ref{secconcl}.

\section{Observational Data}\label{secdata}

The SDSS and its extensions have acquired $ugriz$ photometry for several
hundred million stars; the most recent public release is DR8 \citep{aih11}.
The Sloan Extension for Galactic Understanding and Exploration
\citep[SEGUE;][]{yan09}, one of three sub-surveys that together formed
SDSS-II, extended the $ugriz$ imaging footprint of SDSS-I 
\citep{fuk96,gun98,yor00,gun06,sto02,aba03,aba04,aba05,aba09,pie03,ade06,ade07,ade08}
by approximately 3500 deg$^{2}$, and also obtained $R$ $\simeq$ 2000
spectroscopy for approximately 240,000 stars over a wavelength range of
3800$-$9200\,{\AA}. This included spectra for a collection of Galactic
globular and open clusters, which served as calibrators for the \teff,
\logg, and \feh ~scales for all stars observed by SDSS/SEGUE, as processed
by the SEGUE Stellar Parameter Pipeline \citep[SSPP;][]{lee08a,lee08b,
all08}. Tables \ref{tabclusterphotprops} and \ref{tabclusterspecprops} list
the photometric, spectroscopic, and physical properties of the eight GCs in
our sample. The SSPP produces estimates of \teff, \logg, \feh, and radial
velocities (RVs), along with the equivalent widths and/or line indices for
85 atomic and molecular absorption lines, by processing the calibrated
spectra generated by the standard SDSS spectroscopic reduction pipeline
\citep{sto02}. See \citet{lee08a} for a detailed discussion of the
approaches used by the SSPP; \citet{smo11a} provides details on the most
recent updates to this pipeline, along with additional validations.

\begin{figure}
\centering
\plotone{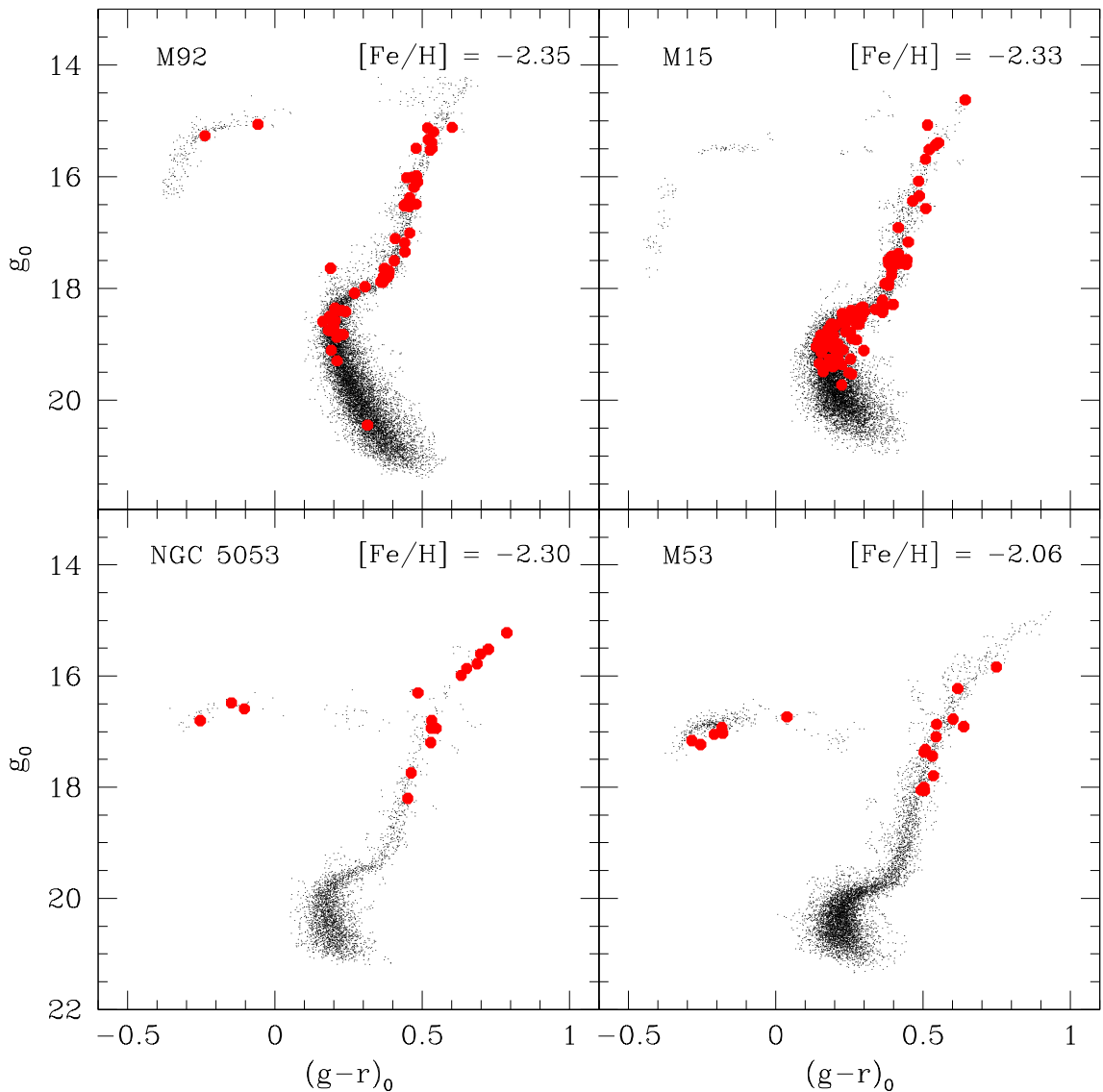}
\caption{Color-magnitude diagrams for the four globular clusters in our
sample with [Fe/H] $< -2.0$: M92, M15, 
NGC 5053, and M53.  The black points represent photometric data for likely
cluster members that passed the tidal radius and CMD mask algorithm cuts.
The red points correspond to spectroscopic data for our selected true cluster
members.}\label{figfirst4cmds}
\end{figure}

\begin{figure}
\centering
\plotone{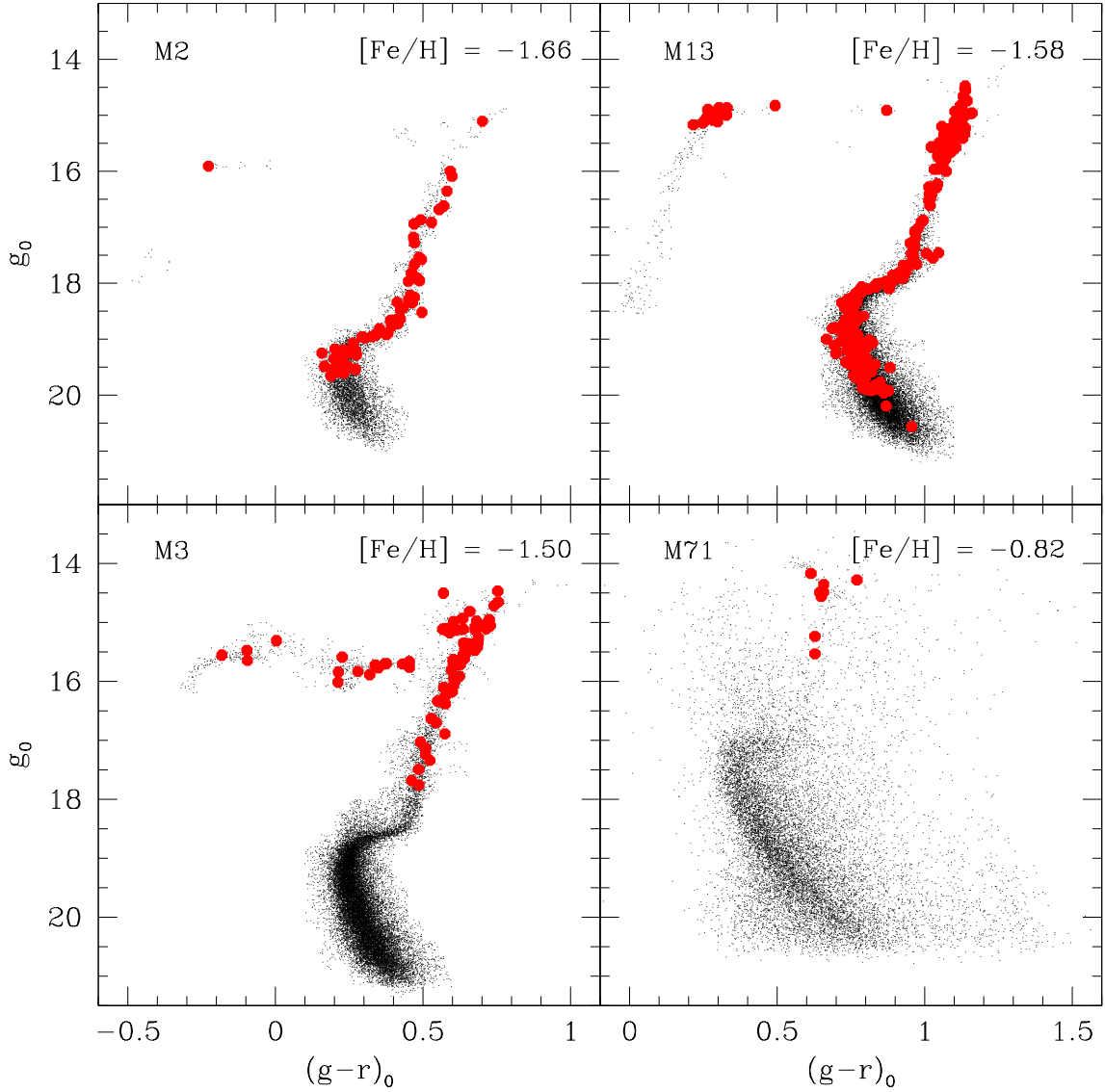}
\caption{Same as Figure \ref{figfirst4cmds}, but for the four globular 
clusters in our sample with [Fe/H] $> -2.0$: M2, M13, M3, and M71.
Membership selection for M71 was slightly different, as described in
Section \ref{secdata}, thus more photometric data is present in the CMD for
this cluster. The black points for this cluster do not all
represent likely cluster members.}\label{figsecond4cmds}
\end{figure}

Membership selection for the clusters is based on the color-magnitude
diagram (CMD) mask algorithm described by \citet{gri95}. Details on the
application of this method to our specific clusters are described by
\citet{lee08b} and \citet{smo11a}, and will only be briefly summarized here.
The procedure involves a series of cuts, reducing the overall sample to
include only those stars for which one can reasonably claim true
membership. First, all stars within the tidal radius of the GC are
selected. Stars with available spectra but with $\langle \rm{S/N}\rangle <
10$ (averaged over the entire spectrum), or that lacked estimates of \feh\
or RV, are excluded. A CMD is then constructed of the remaining stars, along
with a CMD of stars in a concentric annulus designated to represent the
field. A measure of the effective signal-to-noise in regions of the CMD is
obtained, where the ``signal'' in this case constitutes those stars within
the tidal radius and the ``noise'' constitutes those stars within the field
region. Cluster-region stars within segments of the CMD above a threshold
signal-to-noise are then selected. Finally, Gaussian fits to the highest
peaks in the \feh\ and RV distributions of those stars (expected to
represent the cluster) are obtained, and stars within $2\sigma$ of the mean
in both \feh\ and RV are considered true member stars. This procedure
results in the following numbers of true member stars: M92 (58), M15 (98),
NGC~5053 (16), M53 (19), M2 (71), M13 (293), M3 (77), and M71 (8). Figures
\ref{figfirst4cmds} and \ref{figsecond4cmds} show the final CMDs for these
eight globular clusters.  

Membership selection for M71 was complicated due to difficulties
encountered with the photometry values available for this cluster at the
time of our analysis \citep[see][]{an08,smo11a}. This made the CMD mask
algorithm less reliable for selecting likely spectroscopic members.
Therefore, stars inside the tidal radius were selected and passed on to the
final \feh ~and RV cuts, with those stars that had questionable photometry
excluded from consideration. As a result, only eight stars with available
spectroscopy made it through the final cut for this cluster.

\section{CN Bandstrength Distribution}\label{seccn}

A common approach for investigation of the star-to-star light element
abundance variations within GCs is measurement of the 3883\,{\AA} CN
molecular absorption band \citep{nor79,nor81a,smi96,har03a,har03b,pan10}.
This measurement does not require high-resolution spectroscopy, making it
ideal for low-resolution spectroscopic surveys such as SDSS. The feature is
typically measured using a spectral index defined as the magnitude
difference between the integrated flux within a wavelength window
containing the absorption band and the integrated flux within a sideband
representing the continuum. However, the precise definition of this
spectral index is often varied according to the luminosity class of the
stars under consideration, due to the presence of other
temperature-dependent absorption lines, such as H$_{\zeta}$ at 3889\,{\AA},
that can potentially interfere with the adopted continuum window. In this
section, we describe the CN spectral indices used for each region of the
CMD and their observed distributions.

\subsection{CN Absorption on the Red Giant Branch}\label{subseccnrgb}

\begin{figure}
\plotone{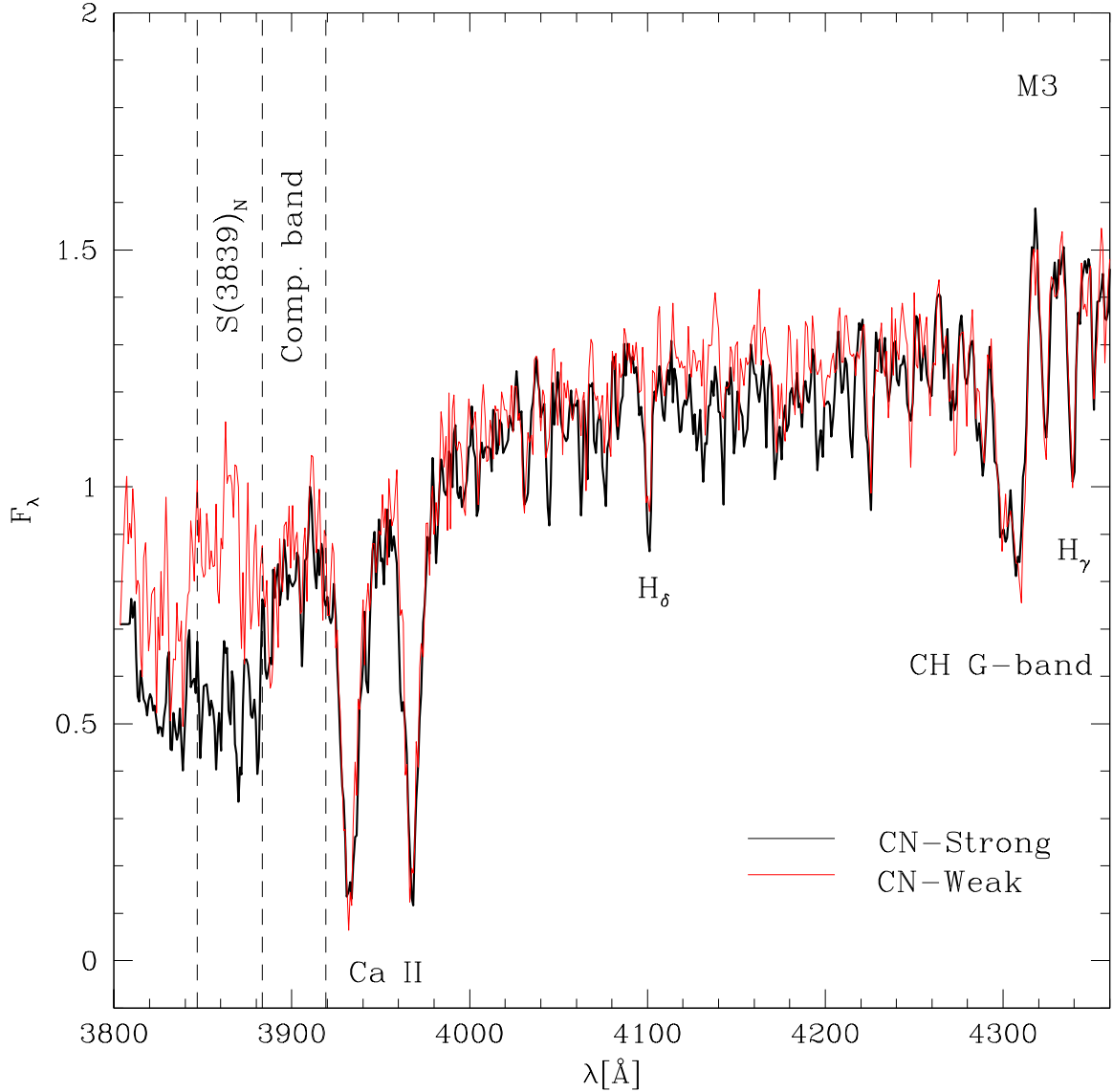}
\caption{Representative blue SDSS spectra of CN-weak (red thin line; fiber 2475-53845-160) and CN-strong 
(black thick line; fiber 2475-53845-489) RGB stars in M3.  The areas
between the dashed vertical lines indicate the portions of the spectrum used in measuring S(3839)$_{N}$.
Other prominent spectral features are labeled.}\label{figcnspeccomp}
\end{figure}

We measured the strength of the CN absorption band at 3883\,{\AA} in RGB
stars using the spectral index S(3839) defined by \citet{nor81a}:

\begin{equation}
S(3839)_{\rm N} = -2.5 \, \rm{log}\frac{\int_{3846}^{3883} I_{\lambda}\rm{d}\lambda}{\int_{3883}^{3916} I_{\lambda}\rm{d}\lambda}, \label{eqcnnor81}
\end{equation}

\noindent where I$_{\lambda}$ is the measured intensity, and the subscript $N$
indicates it is from the \citet{nor81a} definition. Figure
\ref{figcnspeccomp} shows the blue regions of SDSS spectra for two RGB
stars in M3. The line-band and comparison-band windows are indicated. These
two stars were selected because they have similar effective temperatures
and apparent $g$-band magnitudes (indicating similar luminosities on the
RGB), as well as nearly identical Ca~II and CH G-band strengths (indicating
similar metallicities and carbon abundances). Despite these similarities,
they exhibit clear differences in their CN 3883\,{\AA} absorption
strengths.

\begin{figure}
\centering
\plotone{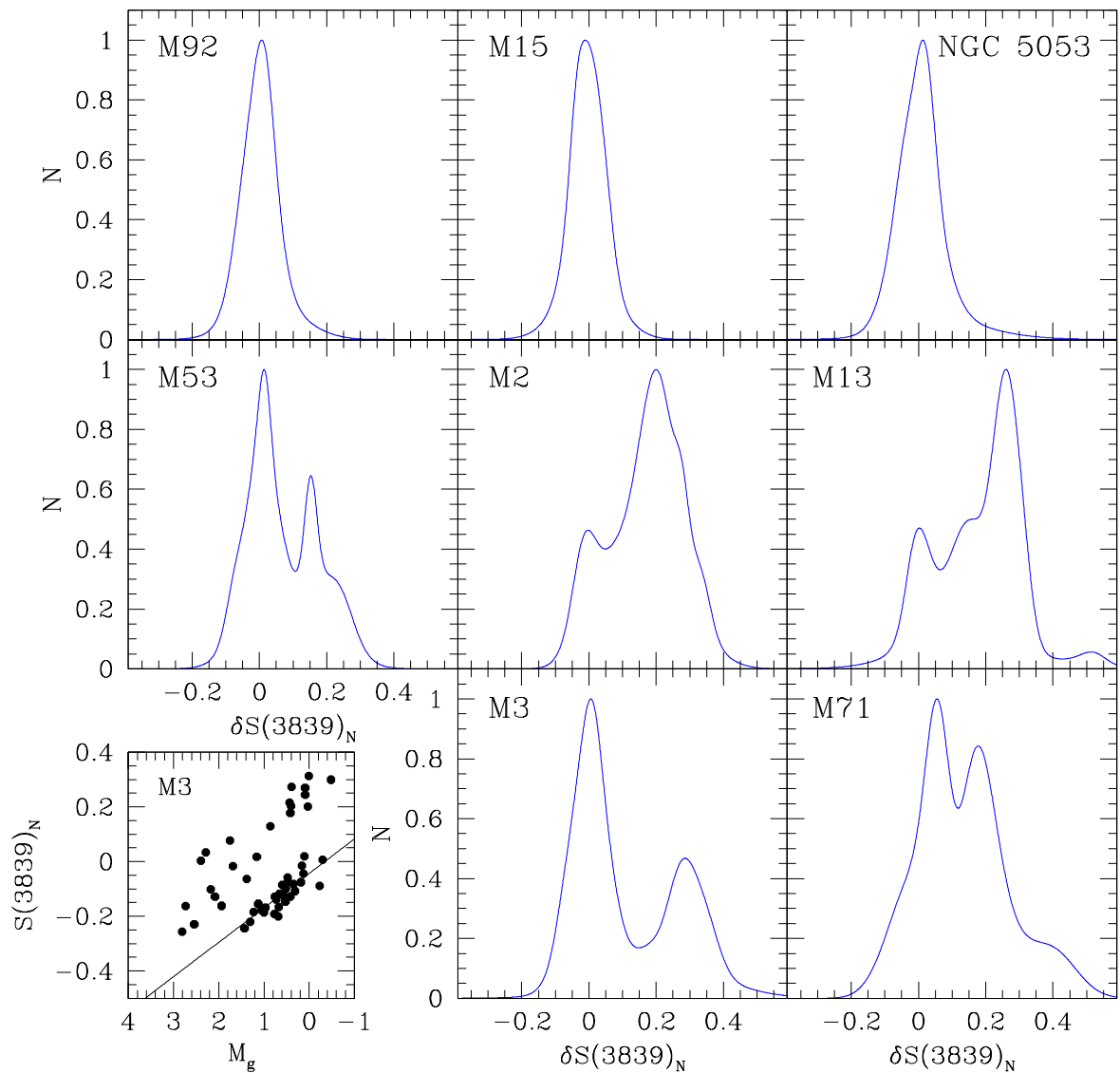}
\caption{Generalized histograms for the $\delta$S(3839)$_{\rm N}$ distributions of RGB stars within
each globular cluster.  The bottom-left panel shows an example of the way 
$\delta$S(3839)$_{\rm N}$ was determined, as described in Section \ref{seccn}, where the
baseline against which $\delta$S(3839)$_{\rm N}$ was measured is shown as a solid line.}
\label{figdcngenhist}
\end{figure}

The formation efficiency of the CN molecule is temperature dependent, where
cooler effective temperatures allow increased molecular formation. When one
looks at a population of stars, one therefore sees an increased ability for
molecular formation moving up the CMD. While the majority of MS stars
(aside from the coolest ones) have effective temperatures too high for
significant molecular formation, the ability for this molecule to form on
the SGB and RGB increases with luminosity as the star expands and its
surface temperature drops. The result of this is that, all things being equal, 
we expect to see increased CN
absorption on the RGB when compared to the SGB and this effect must be
accounted for in the analysis prior to inferring any abundance differences.
Furthermore, for clusters of moderate metallicity, when the CN aborption
strengths are plotted as a function of luminosity or temperature, two
groups generally appear -- one CN-weak (sometimes referred to as
CN-normal), the other CN-strong (enriched). A linear relationship is then
fit to the CN-weak locus, and the vertical difference in S(3839)$_{\rm N}$
between each point and the baseline is measured, as illustrated for M3 in
the bottom-left panel of Figure
\ref{figdcngenhist}. This vertical difference is denoted as $\delta$S(3839)
$_{\rm N}$, and is taken to be a temperature-corrected measure of CN
absorption. The other panels in this figure are generalized histograms of
this temperature-corrected index for our sample of clusters, discussed in
detail below.  The raw and corrected values are listed for each cluster in 
Table \ref{tabcnch}.

\begin{figure}
\centering
\plotone{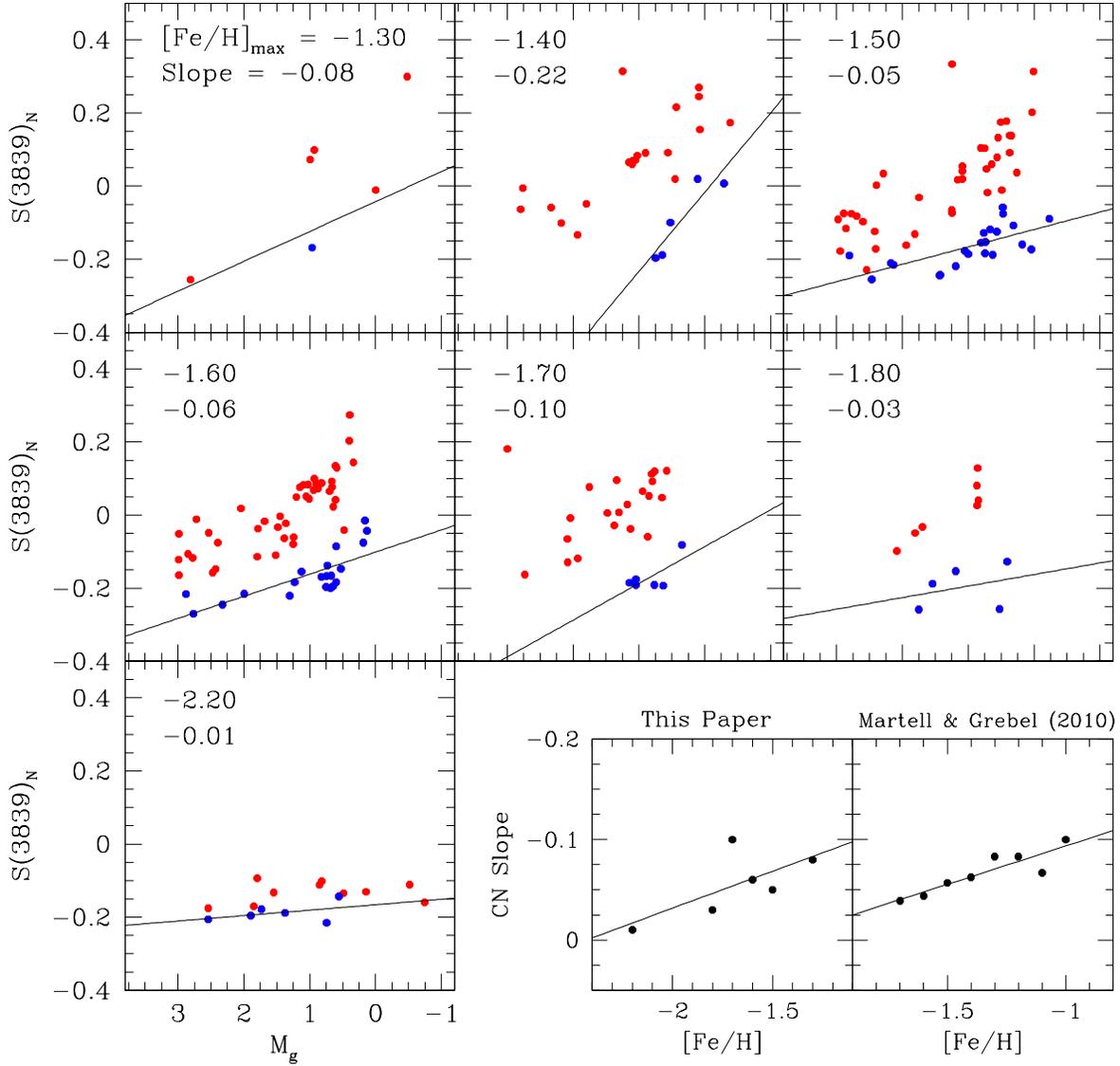}
\caption{Raw S(3839)$_{\rm N}$ values versus absolute $g$-magnitude for seven
0.1-dex-wide metallicity bins for the cluster RGB stars in our sample.
The top number in the upper-left corner of each panel indicates the maximum
metallicity for each bin, while the bottom number indicates the slope of
the fit to the CN-strong locus. Red points indicate stars that were identified
as CN-strong within their respective clusters, while blue points indicate stars
identified as CN-weak.}\label{figcnfehbin}
\end{figure}

The slope of the relationship between CN bandstrength and luminosity is
metallicity dependent, so each cluster must be corrected individually prior
to constructing comparisons across the sample. Figure \ref{figcnfehbin}
shows this relationship between the CN slope and
\feh, obtained by dividing our entire sample into 0.1-dex wide metallicity
bins and fitting a line to the CN-weak locus of each bin. Note the trend of
decreasing CN slope with decreasing \feh, which is similar to the trend for
field giants from DR7 reported by \citet{mar10}. When the slope of each
panel in their Figure 6 is plotted as a function of metallicity, we obtain
a linear relationship for field giants:

\begin{equation}
\rm{CN ~Slope}_{field} = - 0.17 - 0.08\rm{[Fe/H]}. \label{eqcnslfield}
\end{equation}

\noindent The expression obtained for our sample of cluster RGB stars is:

\begin{equation}
\rm{CN ~Slope}_{cluster} = -0.18 - 0.07\rm{[Fe/H],}  \label{eqcnslclust}
\end{equation}

\noindent and is statistically equivalent to the field giant relationship
($\sigma_{\rm{slope}} = 0.04, \sigma_{\rm{intercept}} = 0.07$). The data
point corresponding to \feh = $-1.40$ has been omitted from this linear fit
as an outlier.
% that was poorly constrained by the relative lack of CN-weak
%stars in this bin.  

It is difficult to draw a direct comparison between the two samples due to
their differing mass functions. The cluster giants are all exclusively old
and span a relatively small range in mass at any given value of $M_g$,
whereas the field giants from the \citet{mar10} sample potentially span a
much broader range of mass and age.\footnote{Figure 6 in \citet{mar10}
plots CN versus absolute $M_r$, whereas we use absolute $M_g$. Because
these are all RGB stars with the same $(g-r)_0 \sim 0.5$, converting $M_r$
to $M_g$ should only produce an essentially uniform shift of all stars in
each plot, not affecting the slopes significantly.} Similarities between
these relationships may hint at a common origin \citep[see][for further
discussion]{mar10}.

%Relation between [Fe/H] and size of the CN gap (similar to mar10 fig. 9, but quantify any existing trend)?

\begin{figure}
\centering
\plotone{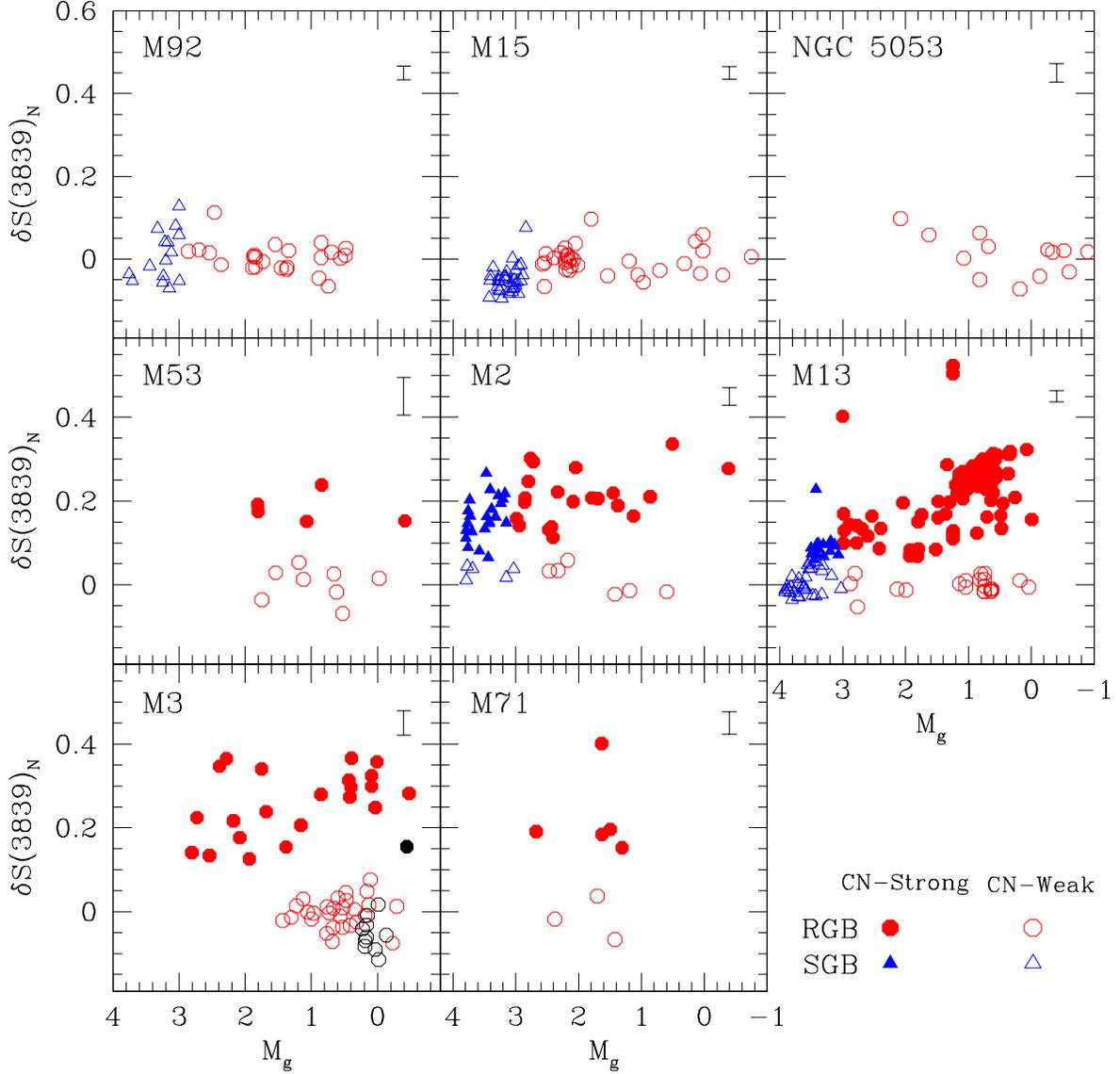}
\caption{The distributions of $\delta$S(3839)$_{\rm N}$ as a function of absolute $g$-magnitude 
for RGB and SGB stars, using the CN index definition from 
\citet{nor81a}.  Blue triangles represent SGB stars and red circles indicate RGB stars.
CN strong stars are shown using filled symbols, while the open symbols represent
CN weak stars.  The black points in M3 are AGB stars.  Typical uncertainties are indicated
by the vertical line in the upper right corner.  Clusters are arranged in order of 
increasing [Fe/H].}\label{figdcndist}
\end{figure}

Figure \ref{figdcndist} shows the distribution of $\delta$S(3839)$_{\rm N}$
as a function of absolute $g$-magnitude. Values for both RGB and SGB stars
are shown, calculated using the CN index definition of \citet{nor81a}. Blue
triangles represent SGB stars, while red circles represent RGB stars, with
filled and open symbols indicating CN-strong and CN-weak stars,
respectively. While a small amount of scatter exists for the most
metal-poor clusters (M92, M15, and NGC~5053), no separation that would
indicate a bimodal distribution is obvious, consistent with Figure
\ref{figdcngenhist}. For the remaining five clusters at higher metallicity,
starting with M53 at a metallicity of \feh = $-$2.06, two distinct
populations of stars in $\delta$S(3839)$_{\rm N}$-space are apparent. 

For M3, a small number of possible AGB stars are noted in the figure with
black symbols. The majority of these stars appear CN-weak, with only one
CN-strong AGB star. This result is similar to the observations reported by
\citet{cam10}, who found that, in a sample of nine Galactic GCs, all showed
either a total lack of CN-strong AGB stars or a significant depletion of
CN-strong AGB stars compared to those present on the RGB. These authors
noted that no current explanation exists in standard stellar evolution
theory as to why stars on the AGB should have reduced CN abundances
compared to the RGB, particularly because the low effective temperatures
should be suitable for similarly efficient molecular CN formation. In
principle, increased mixing both on the RGB and at the beginning of AGB
ascent should contribute more N (and thus stronger CN) to the stellar
envelope, which should be apparent in surface abundance measurements. Such
a discrepancy has been noted for a long time; two possible explanations
were proposed by \citet{nor81a}. First, if two chemically distinct
populations in the cluster existed after star formation ceased, one of
which was helium-rich and evolved to populate the blue end of the
horizontal branch (HB), but never ascended to the AGB, this might lead to
the deficiency of CN-strong stars. The second explanation hypothesized that
increased mixing in some stars produced increased CN abundances, but also
led to increased mass loss at the RGB tip, producing stars populating the
blue end of the HB that never ascended the AGB. The problem remains
unsolved, and requires additional work.

Adopting the corrected values obtained from the above procedure, we
produced a generalized histogram of the $\delta$S(3839)$_{\rm N}$
distribution for each cluster, shown in Figure \ref{figdcngenhist}. This
was accomplished by representing each point as a Gaussian, centered on
$\delta$S(3839)$_{\rm N}$ with a FWHM equal to the uncertainty of that
particular $\delta$S(3839)$_{\rm N}$ measurement, and then adding the
individual Gaussians together. The uncertainty for each S(3839)$_{\rm N}$
measurement was determined as in \citet{mar10}, using a Monte Carlo
approach. Each pixel in the error vector produced by the SDSS spectroscopic
reduction pipeline was multiplied by a factor between 0 and 1, drawn from a
normalized distribution. This new vector was then added to the data vector
and the indices were remeasured. This process was repeated 100 times and
the standard deviation was taken to be the uncertainty. Naturally, the
resultant uncertainties are related to the S/N of the spectra;
uncertainties for the high-S/N spectra were much lower than the typical
uncertainties found in the literature. For this reason, recent studies have
sometimes smoothed their histograms to make them more directly comparable
to past studies. Smoothing the histogram also helps eliminate any
artificial substructure in the distribution created by small number
statistics, while additionally accounting for unidentified sources of
uncertainty.

In Figure \ref{figdcngenhist}, the clusters are arranged in order of
increasing metallicity, from left to right, top to bottom, and represent
the $\delta$S(3839)$_{\rm N}$ distribution on the RGB for each cluster.
Many of these clusters have been studied previously, so comparisons can be
made with our present observations. \citet{sun81} reported a bimodal
distribution in CN indices in M3 and M13 on the upper RGB (stars more
luminous than the HB). This observation for M3 was confirmed on the upper
RGB by \citet{smi96} and \citet{lee99}, and on the lower RGB of M3 by
\citet{nor84}. \citet{smi96} also report CN bimodality among M13 RGB stars,
and the proportions of CN-strong stars we observe in these two clusters
agree well with those reported by \citet{sun81}. M2 was studied by
\citet{smi90} and shown to have a bimodal CN distribution, also matching in
proportion to that seen in our data.

Studies of M71 by \citet{smi82a}, \citet{lee05}, and \citet{alv08} all
report CN bimodality on the RGB. Evidence for bimodality is found in our
data as well, at the same level as observed for 47~Tuc \citep{nor79}, which
is of comparable metallicity to M71. It is interesting that \citet{alv08}
have claimed the existence of CN-strong AGB stars in their sample. As
mentioned above, nearly all CN observations of AGB stars have demonstrated
a depletion in CN, which makes their observation unique. It is possible
that M71 is so metal-rich compared to other GCs studied to date that this
encourages additional CN enrichment on the AGB, in spite of whatever
mechanism might be causing the depletions in other cluster AGB stars.
However, if this were the case, one might expect some manifestation on the
RGB as well, in the form of a higher ratio of CN-strong to CN-weak stars.
Yet, in the \citet{alv08} sample the ratio is only $\sim 0.3$. Further
investigation of AGB stars in M71 would be of interest to determine whether
their observations are representative of the cluster. Note that
\citet{mal78} reported a large fraction of CN-strong AGB stars in 47~Tuc,
so we might expect to find a similar fraction in M71.

\begin{figure}
\centering
\plotone{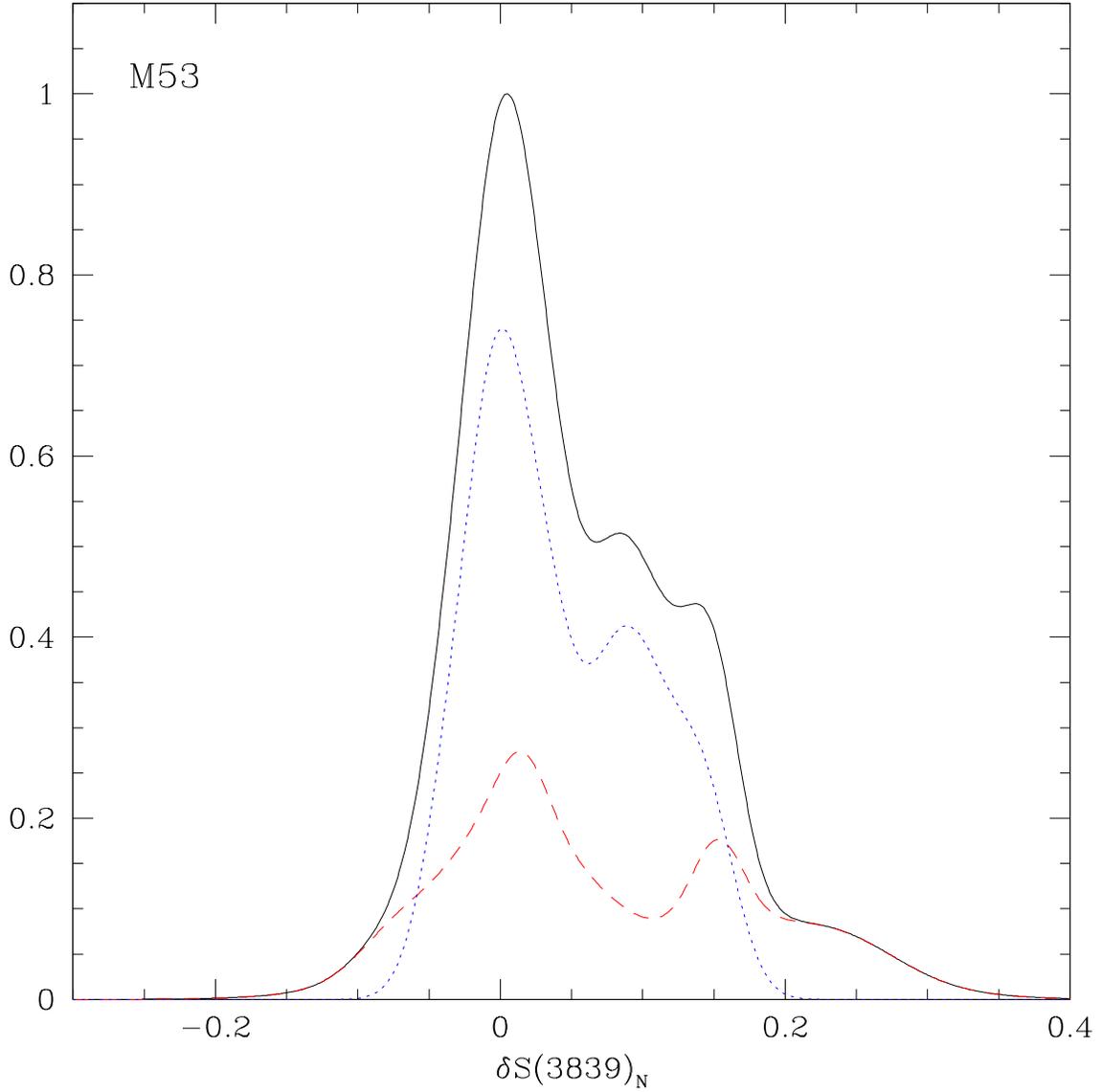}
\caption{Generalized histogram for the combined (solid black line) 
$\delta$S(3839)$_{\rm N}$ data sets for RGB stars in M53 from our sample
(red dashed line) and that of \citet{mar08a} (blue dotted line). This 
distribution suggests that while there is a prominent CN-weak group, a 
group of CN-strong stars also exists in this cluster.}\label{figm53dcngenhist}
\end{figure}

Moving to the \feh ~$\leq -2.0$ regime, M53 has not been extensively
studied. \citet{mar08a} reported a broad but not strongly bimodal
distribution of CN absorption strengths in their sample of upper RGB stars
(brighter than the RGB bump), where deep mixing is expected to have altered
the stellar surface abundances. Their Figure 6 shows a generalized
histogram for M53 that is similar to ours in Figure \ref{figdcngenhist},
but theirs appears slightly narrower and smoother (though in fact our
histogram has been smoothed by a larger factor than their distribution).
When our $\delta$S(3839)$_{\rm N}$ values are combined with theirs,
producing a sample spanning nearly the entire RGB, a KMM test
\citep{ash94} indicates that the hypothesis that the observations are drawn
from a single Gaussian parent population can be rejected at high
statistical confidence ($p = 0.05$). The generalized histogram for this
combined data set is shown in Figure \ref{figm53dcngenhist}, where the
individual data sets are also indicated. The
distribution clearly indicates the presence of a CN-strong component, with
a ratio of CN-strong to CN-weak stars of 0.61, suggesting a population of
CN-strong stars in the cluster with a range of enrichment levels.

We now consider the other very metal-poor clusters in our sample: M92, M15,
and NGC~5053. \citet{car82} studied carbon and nitrogen abundances in M92
from the SGB to the AGB, and reported a general decrease in C abundances
moving to higher luminosities, but no correlation or anticorrelation
between the C and N abundances. Instead, they determined that the total
number of C+N atoms varies from star to star within the cluster, an
observation at odds with predictions that the sum ought to remain constant.
Similar conclusions were reached by \citet{tre83} regarding M15 -- C
abundance drops as one moves up the RGB, but N abundance remains on average
the same, although with some uncorrelated star-to-star variations. A more
recent study of M15 by \citet{lee00} also failed to detect reliable
evidence of CN bimodality, although they confirmed the existence of a very
small number of CN-enriched stars, also reported by \citet{lan92}.  

\begin{figure}
\centering
\plotone{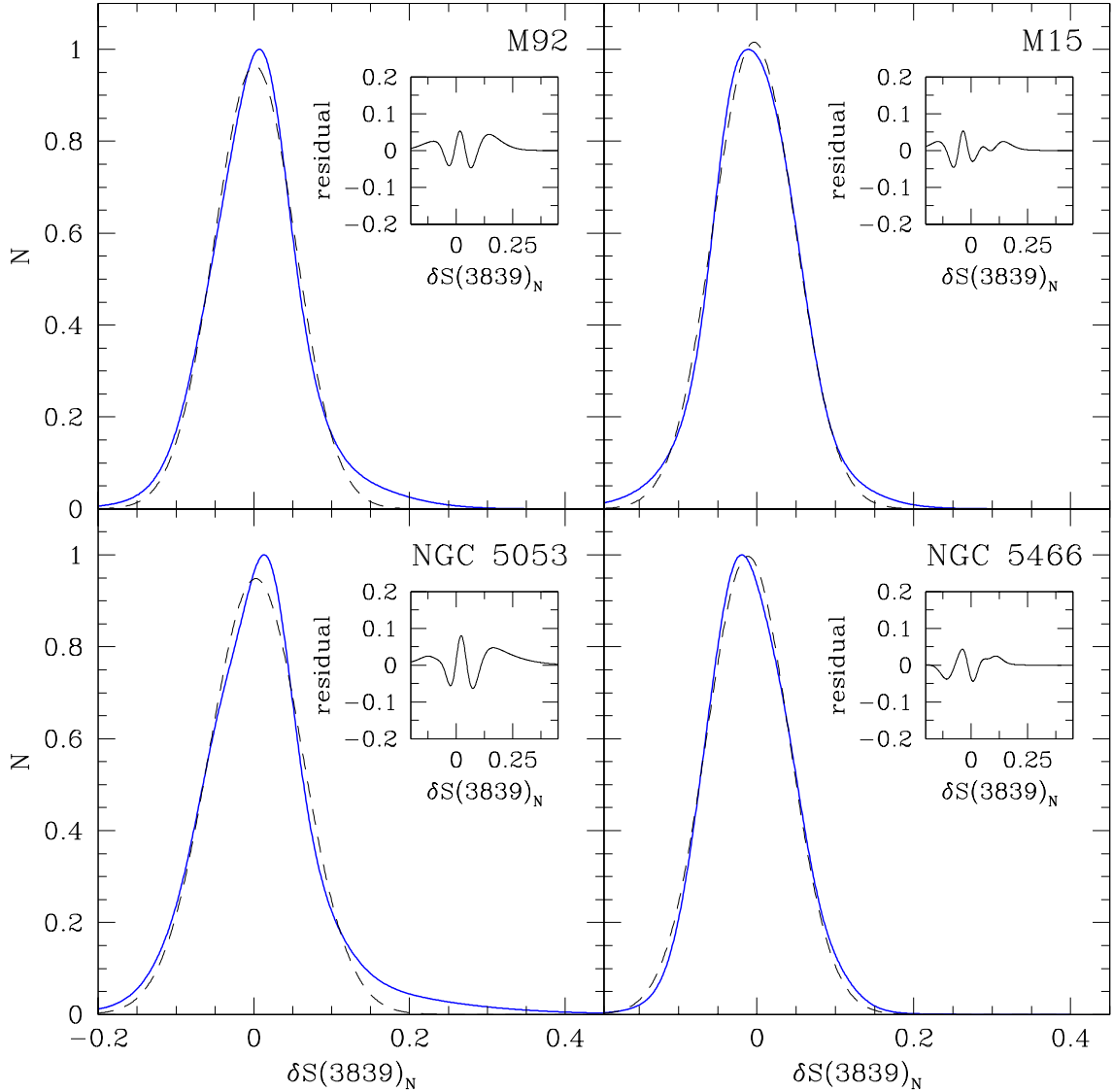}
\caption{Generalized histograms for RGB stars in the very metal-poor
clusters M92, M15, and NGC 5053, along with NGC 5466 ([Fe/H] $= -2.2$),
taken from \citet{she10}. The blue solid line is the data, while the black
dashed line represents the single Gaussian that best fit the distribution.
The inset shows the residual between this best-fit curve and the data; the
double-peaked nature of the residual may suggest that the data may be best
represented by two overlapping Gaussian
distributions.}\label{figm92m15n5053n5466rgbgenhist}
\end{figure}

A recent study by \citet{she10} of NGC 5466 (\feh ~$\simeq -2.2$) suggested
the possible presence of two CN groups, with a small mean separation of
only 0.055. They noted that the generalized histogram of their RGB stars
was not well-described by a single Gaussian fit. In a similar fashion, we
examined the generalized histograms for RGB stars in the very metal-poor
clusters M92, M15, and NGC~5053. Figure \ref{figm92m15n5053n5466rgbgenhist}
shows the histograms for these clusters, as well as that of NGC~5466 from
\citet{she10}, fit with a single Gaussian; residuals are plotted in the
insets in each panel. The residuals from the fits to our three very
metal-poor clusters are clearly larger than those of NGC~5466, suggesting
that our data also may not be well-described by a single Gaussian population.
However, a KMM test for each cluster (including NGC~5466) cannot reject the
null hypothesis that a single population well-describes the observed data,
indicating that hints of the non-Gaussian distributions in our data may
simply be due to small-$n$ statistics. If these clusters possess multiple
CN behaviors, they may not be discernible within the present measurement
uncertainties. Due to the fact that double-metal molecules like CN are
particularly difficult to observe at low \feh, this problem could be solved
by measuring individual element abundances rather than molecular
bandstrengths.

\subsection{CN Absorption on the Subgiant Branch and Main Sequence}\label{subseccnms}

Perhaps the most intriguing observations of the CN bimodality phenomenon
are to be found among stars in the relatively unevolved regions of cluster
CMDs. Contrary to predictions of standard stellar evolutionary models,
significant variations in CN bandstrengths have been reported for stars
prior to their undergoing first dredge-up \citep[eg.][]{bri02,coh02}, even
down to the main sequence in 47~Tuc \citep{can98,har03a}. However, searches
within other cluster MS stars have produced mixed results. \citet{coh99a}
reported no significant CN variation for MS/MSTO stars belonging to M13,
although the CN features in their spectra were shown to be too weak for
reliable measurement \citep{bri01}. Carbon and nitrogen abundance analyses
of these stars by \citet{bri04} showed that this was likely due to the fact
that there is very little change in bandstrength for a given change in
abundance at luminosities near the turnoff, where effective temperatures
are relatively high compared to MS and giant stars (see their Figure 2).
Main sequence stars in M71 have been claimed to exhibit CN bimodality at a
level larger than the measurement uncertainty, as well as an
anticorrelation between CN and CH \citep{coh99b}. Follow-up analysis of
this data further showed that the variation is at the same level as that
observed for RGB stars in that cluster, leading the authors to claim that
no significant mixing is occurring on the RGB (although this could also
simply mean that first dredge-up did not significantly affect the surface
carbon and nitrogen abundances), and that the abundance variations were in
place at the time the stars formed \citep{bri01}. In their sample of eight
GCs, \citet{kay08} found no statistically significant variation in CN
abundance for stars on the MS and SGB, but they again attributed that to
low S/N spectra producing relatively large measurement uncertainties.
Finally, \citet{pan10} reported CN bimodality for MS stars in four of their
most metal-rich clusters among a sample of 12 clusters. Clearly, minimizing
measurement uncertainty plays a vital role in addressing the question of CN
abundance variations on the MS, and further observations of larger samples
of MS stars are needed for improved statistical certainty. In this
section, we report on the MS/SGB stars observed for four clusters in our
sample.

We measured the strength of the CN absorption band at 3883\,{\AA} on the main
sequence using the spectral index S(3839) defined by \citet{har03a} for MS
stars:

\begin{equation}
S(3839)_{\rm H} = -2.5 \, \rm{log}\frac{\int_{3861}^{3884} I_{\lambda}\rm{d}\lambda}{\int_{3894}^{3910} I_{\lambda}\rm{d}\lambda}. \label{eqcnhar03}
\end{equation}

\begin{figure}
\centering
\plotone{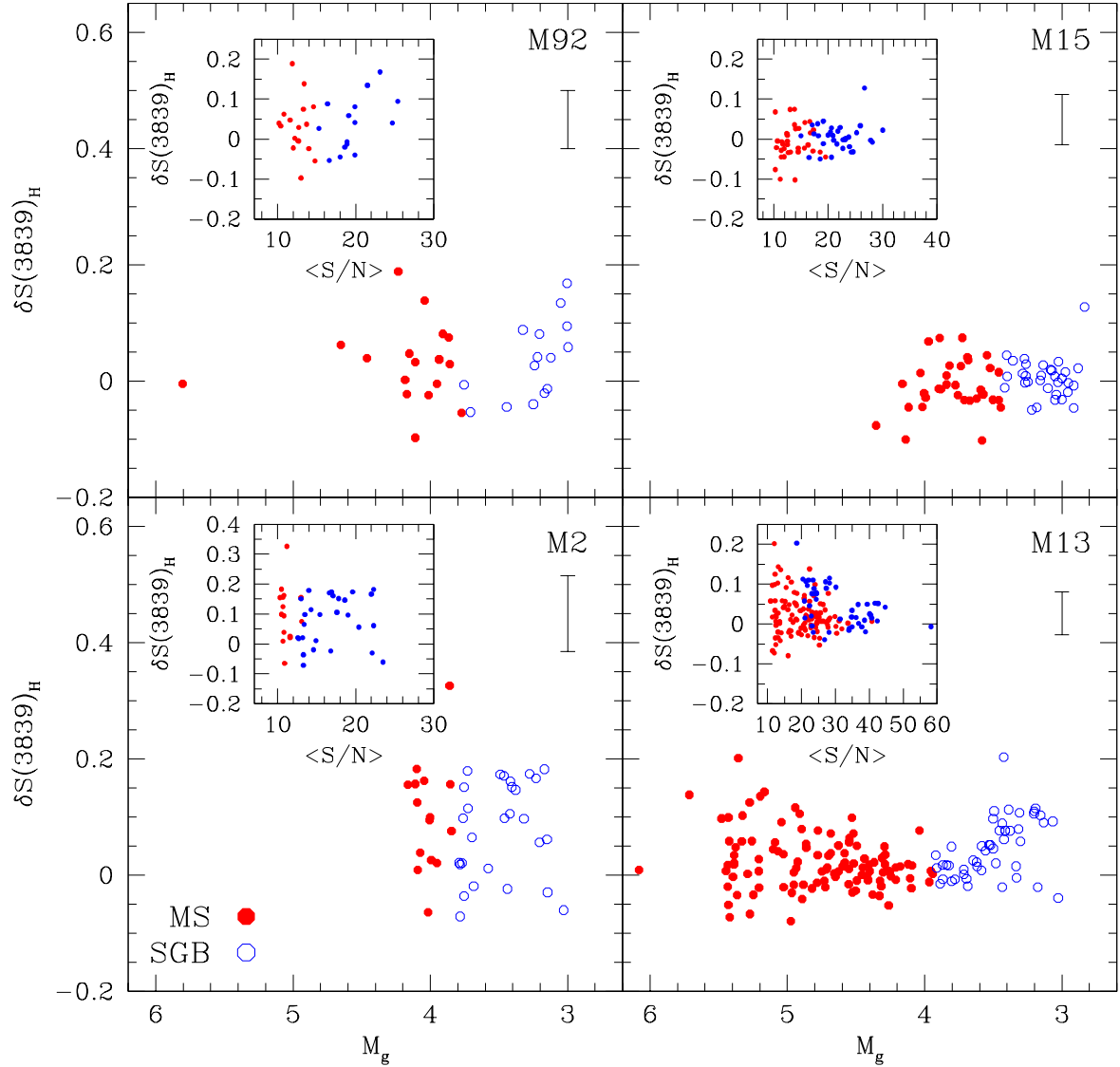}
\caption{Distribution of $\delta$S(3839)$_{\rm H}$ as a function of absolute $g$-magnitude for
MS and SGB stars from clusters with SEGUE spectra on the MS, using the CN index definition from \citet{har03a}.
Blue open circles represent SGB stars and red filled circles represent MS stars.
Plotted as insets for each cluster are the distributions of $\delta$S(3839)$_{\rm H}$ as a 
function of $\langle \rm{S/N}\rangle$.  This is done to demonstrate that the large
scatter of CN absorption strength on the MS may simply be due to low S/N.
Typical uncertainties are indicated by the vertical line in the upper right
corner.  Clusters are arranged in order of increasing [Fe/H].}\label{figdcnsnms}
\end{figure}

\begin{figure}
\centering
\plotone{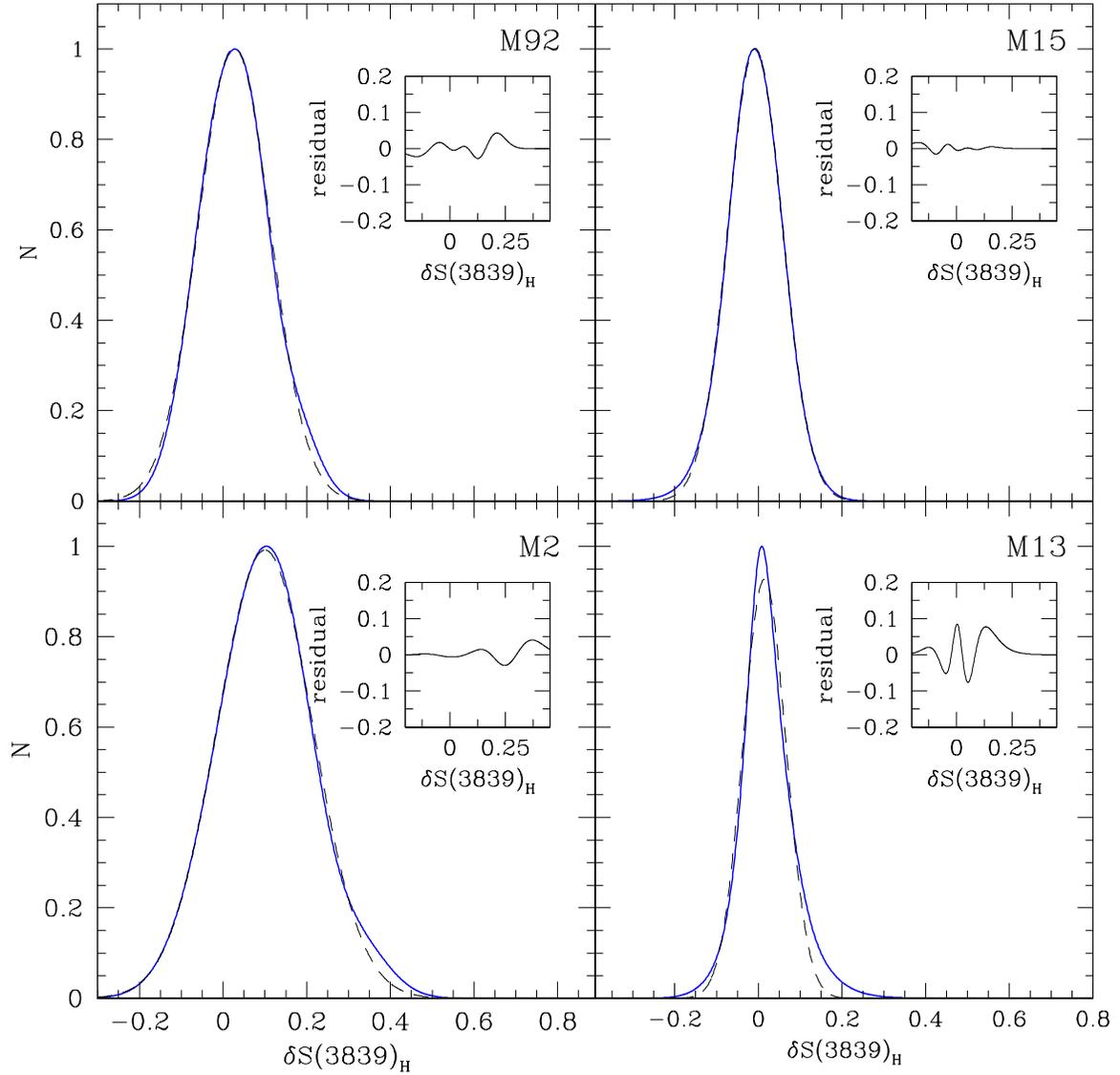}
\caption{Generalized histograms for the $\delta$S(3839)$_{\rm H}$ distributions of MS stars within each globular cluster
for which SEGUE spectra exist on the MS.  $\delta$S(3839)$_{\rm H}$ on the MS is calculated in the same way as for 
RGB stars.  No indication of bimodality is apparent.}\label{figdcngenhistms}
\end{figure}

\noindent Uncertainties and
$\delta$S(3839)$_{\rm H}$ values were calculated the same way as described in
Section \ref{subseccnrgb}. Figure \ref{figdcnsnms} shows the distribution
of $\delta$S(3839)$_{\rm H}$ values for the four clusters that have SEGUE spectra
for stars on the MS -- M92, M15, M2, and M13 (in order of increasing \feh).
A large amount of scatter is apparent, but not when compared to the typical
uncertainty indicated by the error bars shown in the upper right corner of
each panel. Furthermore, inset in each panel is the distribution of
$\delta$S(3839)$_{\rm H}$ as a function of $\langle \rm{S/N}\rangle$, which shows
that the source of this scatter may lie in the relatively low S/N of the
spectra for these faint stars. The decrease in CN strength near M$_g \approx 4$ is
not unexpected, since this corresponds to the turnoff where the effective 
temperatures are the highest (and thus CN molecular formation is at its lowest),
but one would expect that CN bandstrengths should essentially all increase for luminosities
below this point again, as it does for luminosities higher than this point, 
rather than simply increasing in dispersion.  Generalized histograms of the
$\delta$S(3839)$_{\rm H}$ values for these four clusters are shown in Figure
\ref{figdcngenhistms}. No indications of bimodality are seen, suggesting
that when the relatively larger uncertainties are taken into account,
nothing statistically significant stands out. Although the residuals (see inset
panels) are asymmetric and, in the cases of M92 and M13, double-peaked, we
see nothing indicating the presence of two populations of stars. It seems
more likely that the asymmetries in the figures are simply due to finite
sampling from a single Gaussian distribution. A KMM test fails to reject
the hypothesis that these data were drawn from a single Gaussian parent
population. Further observations with smaller uncertainties are needed to
determine whether the observed distributions' asymmetries are due to the
presence of two distinct populations or not.

\begin{figure}
\centering
\plotone{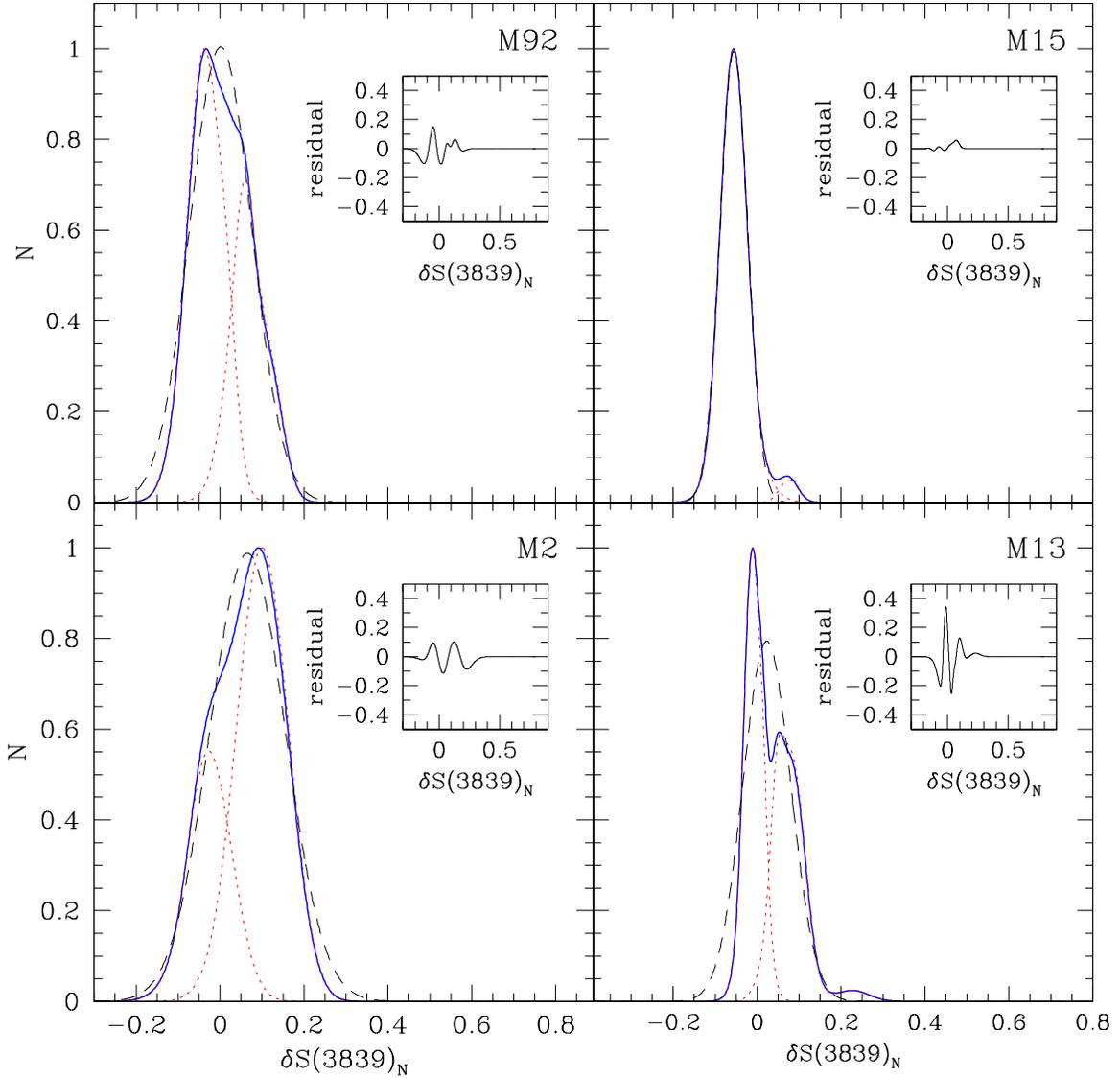}
\caption{Generalized histograms for the $\delta$S(3839)$_{\rm N}$ distributions of SGB stars within each globular cluster
for which SEGUE spectra exist on the SGB. The solid blue curves are the generalized histograms for each cluster.  A simple 
Gaussian fit to the distribution is overplotted as a dashed black line, and the red dotted curves are generalized
histograms for the groups we might presume to be CN-strong and CN-weak, taken separately.  The inset box shows the 
difference between the blue generalized histogram for the entire sample of SGB stars and the simple Gaussian 
fit.  The small bump on the CN-strong side represents one star and cannot be confidently assigned to any presumed
CN-strong group.}\label{figdcngenhistsgbn} 
\end{figure}

The \citet{nor81a} definition for the CN index was used for stars located
on the SGB, although the \citet{har03a} definition is also valid. Figure
\ref{figdcngenhistsgbn} shows the distributions of $\delta$S(3839)$_{\rm N}$
abundances on the SGB for the same four clusters as in Figure
\ref{figdcngenhistms}. The histograms for M92, M2, and M13 appear to
provide evidence of independent CN groups. The solid blue lines and dashed
black lines are as before, while the red dotted curves provide the
generalized histograms for the proposed CN groups on the SGBs of each
cluster. Insets in each panel again show the differences between the data
and the superposed single Gaussian curve. \citet{coh99a} looked for CN
variations on the upper MS/MSTO region of M13 and reported nothing
significant; however it appears that our data does indicate the presence of
CN variation on the SGB of this cluster. These observations indicate that,
for very metal-poor clusters such as M92, as well as for clusters with
moderate metallicity such as M2 and M13, there appear to be signs of
enhanced N enrichment well before the point of first dredge-up. These
issues are explored further below.

\subsection{Hidden Substructure in Generalized Histograms}\label{subsecgenhist}

While generalized histograms are a more natural representation of the
distribution of data than binned histograms, it is important to consider
the impact of any adopted smoothing factors. Smoothing factors are
sometimes used to produce a histogram that more closely resembles those of
past studies by multiplying the uncertainties of each data point by some
appropriate factor. This is also done to account for any sources of
uncertainty that may have been overlooked. One may choose to adopt a
smoothing factor which produces a distribution that is comparable to past
studies, but choosing a smoothing factor that is too high may wash out
important details in the data.

\begin{figure}
\centering
\plotone{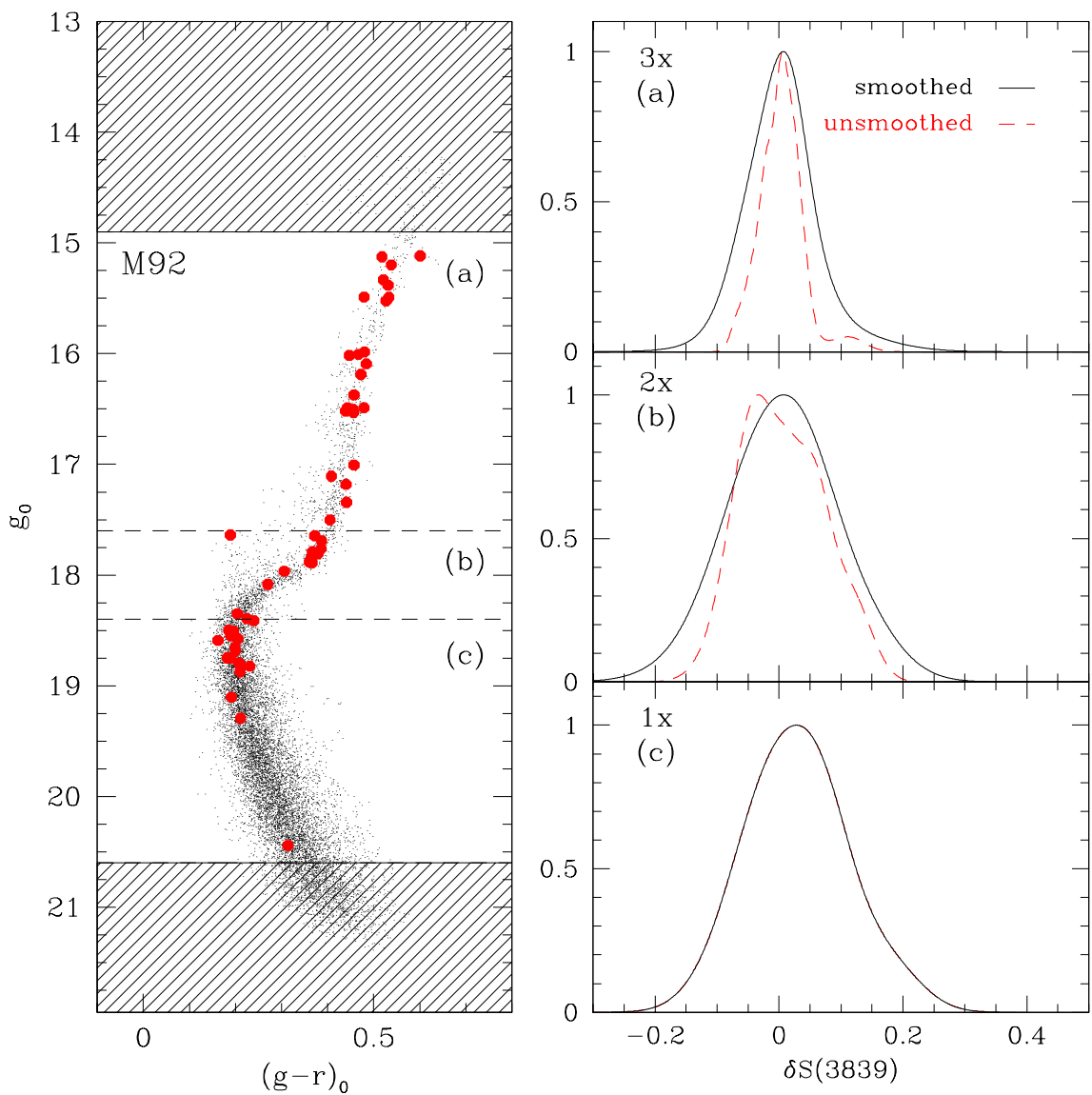}
\caption{Region divisions for the color-magnitude diagram of M92 are shown in the left panel, while the right panel
shows the generalized histograms of the $\delta$S(3839) distribution within each unshaded region.  
%The RGB was divided at 
%approximately the point of first dredge-up, with the other regions representing the SGB, MSTO, and MS.  
The solid black
lines in the histograms represent the smoothed distribution, with the smoothing factor given in the upper-left corner,
while the dashed red lines represent the unsmoothed distribution.  
%In the top-right panel, two smaller inset histograms
%are shown which present the smoothed (black lines) and unsmoothed (red lines) distributions for the two individual groups
%of stars on the upper region of the RGB.  The top and bottom inset histogram correspond to the upper and lower group,
%respectively.  
%While the overall distribution in the uppermost region of the RGB shows no signs of substructure, each
%of the two individual groups do.  
The multipliers listed in the upper left corner of the right-hand set of
panels indicate the amount of smoothing employed (see text).  The HB stars were not included in the analysis.}\label{figm92cmddiv}
\end{figure}

\begin{figure}
\centering
\plotone{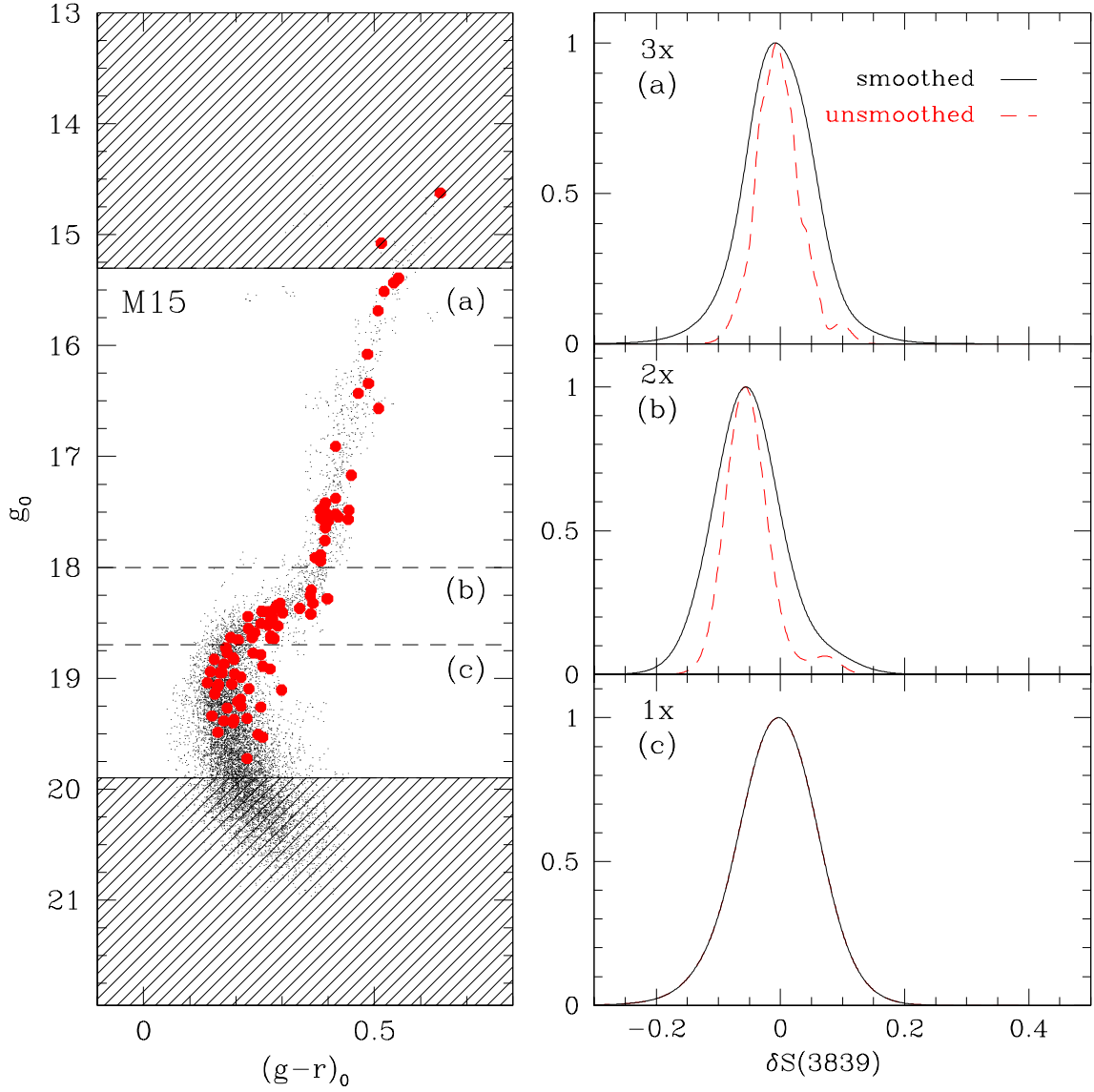}
\caption{Same as Figure \ref{figm92cmddiv} but for M15.}\label{figm15cmddiv}
\end{figure}

\begin{figure}
\centering
\plotone{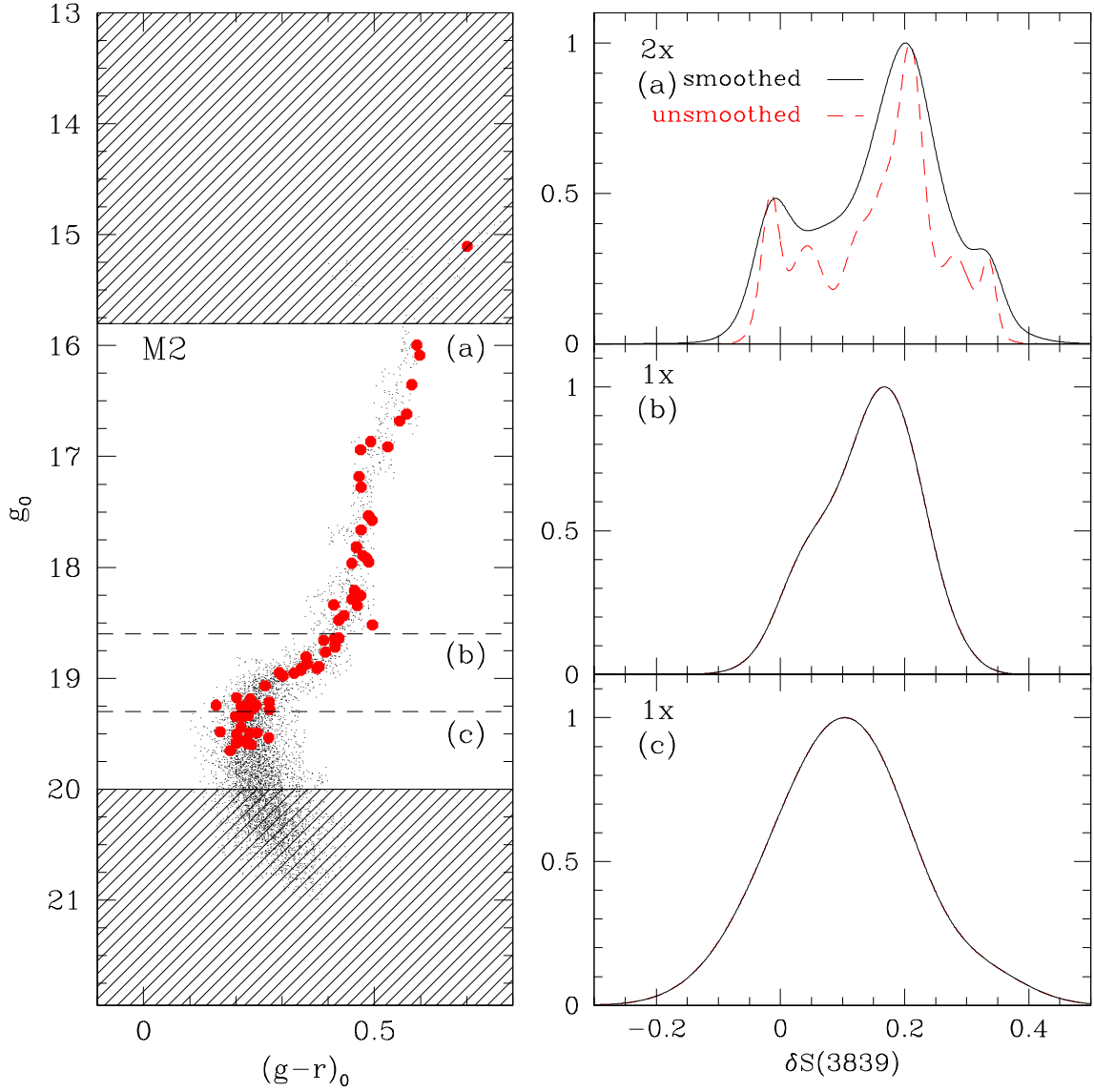}
\caption{Same as Figure \ref{figm92cmddiv} but for M2.}\label{figm02cmddiv}
\end{figure}

\begin{figure}
\centering
\plotone{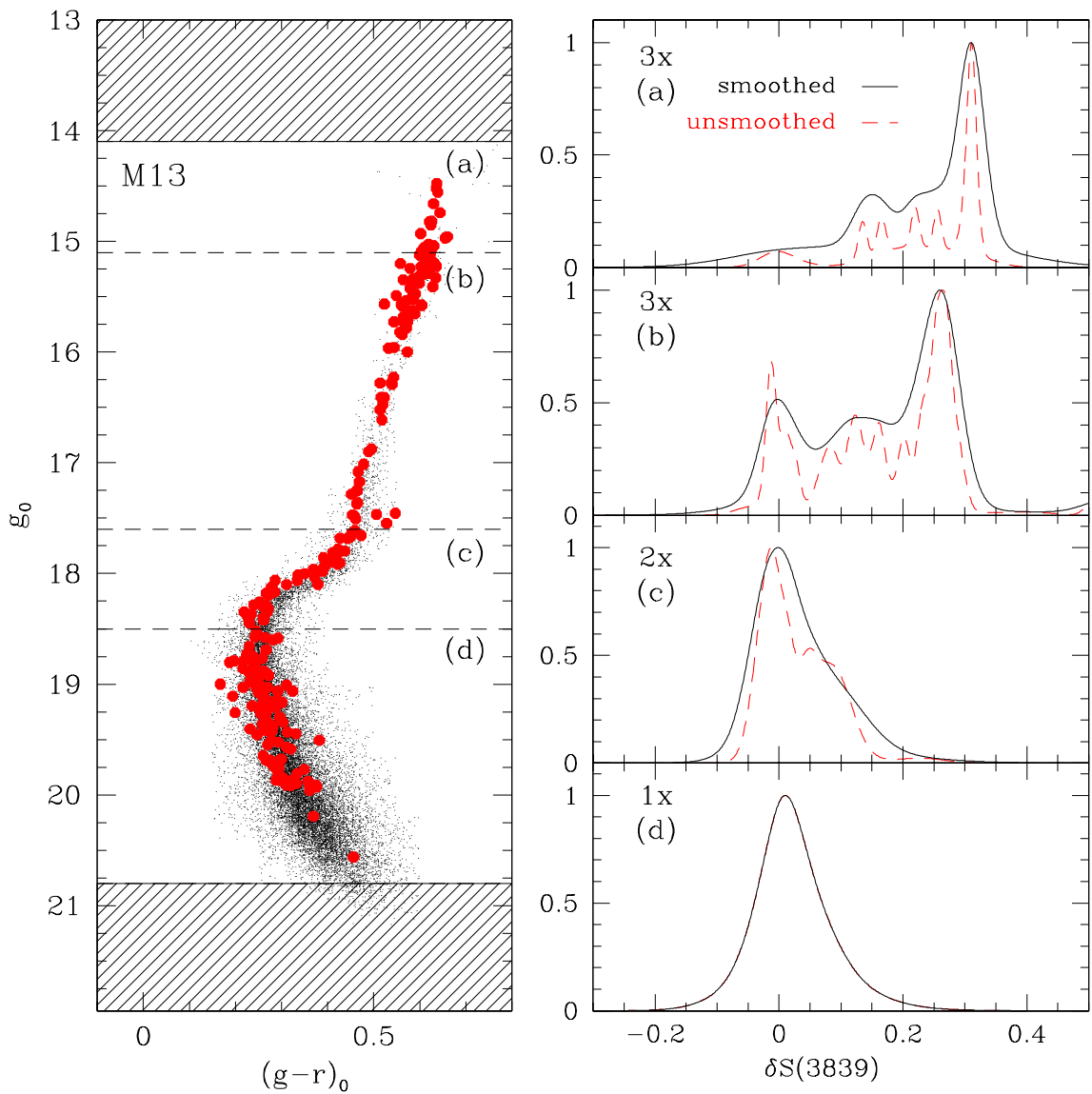}
\caption{Same as Figure \ref{figm92cmddiv} but for M13.  This cluster had data sampled above the
RGB bump, shown in panel (a) of the right-hand set of panels.
More than the other clusters, M13
shows a steady progression of increasing CN richness as one moves up the RGB.  The small bump at $\delta$S(3839) $\approx 0.0$ 
in the top unsmoothed histogram corresponds to an AGB star, identified later as such by its $u-g$ color.}\label{figm13cmddiv}
\end{figure}

To study the potential impact of smoothing on the generalized histograms of
our clusters, we divided up the CMDs for the four GCs with full CMD
coverage into several regions -- RGB above the bump (where the RGB bump was
identified), RGB, SGB/MSTO, and MS. We then looked at the $\delta$S(3839)
indices for stars in each region and produced two generalized histograms,
one smoothed to match the observational uncertainties of previous studies
(solid black line) and one unsmoothed (dashed red line), shown in Figures
\ref{figm92cmddiv} -- \ref{figm13cmddiv}. Because measurement errors from
previous studies are typically $\sim 0.05$, when our typical Monte
Carlo-calculated uncertainties were smaller than this they were amplified
(smoothed) by an integer factor (shown in each panel) to approximate the 
errors from previous studies.  

Naturally, the unsmoothed lines exhibit more potential substructure, but
one must still determine what level of substructure is meaningful. This can
be qualitatively estimated as a function of the number of stars used in the
bin and the relative peak sizes. For example, while the RGB of M92
(Figure \ref{figm92cmddiv}) appears to exhibit substructure in the
unsmoothed histogram, the paucity of stars in this region of the CMD
obviates this claim (the CN-strong peak only has one star). As mentioned
above, for very metal-poor clusters such as M92 and M15, it is expected
that the difference between CN-strong and CN-weak groups should be smaller
when measuring the S(3839) index. At the metallicity of M92, this
difference is expected to be on the order of 0.1, suggesting that if more
data were available we might be able to
determine whether the apparent substructure is real or merely an artifact
of small number statistics. 

From inspection of the RGB of M15 (Figure \ref{figm15cmddiv}), we again see
possible asymmetries in the upper regions that may be associated with
substructure. However, a KMM test for this portion is again unable to
reject a single Gaussian parent population.

At more moderate metallicities, M2 (Figure \ref{figm02cmddiv}) and M13
(Figure \ref{figm13cmddiv}) exhibit very clear signs of CN variation on the
RGB. The uppermost region of M13 is at a luminosity above the RGB bump,
suggesting that the increase in CN is likely due at least in part to the
deep mixing thought to occur at that point. Figure \ref{figm13cmddiv}
depicts the expected appearance of stronger CN bandstrengths as stars
evolve up the RGB, although a group of CN-weak stars always remains
present. It is also worth noting that initial variations first appear on
the SGB, as seen in the right-hand panel (c).

\section{CH Bandstrength Distribution}\label{seccnch}

While investigation of CN bandstrengths can provide some insight into
possible chemical inhomogeneities within cluster stars, CH absorption
strength is also typically considered as a means of distinguishing between
nitrogen abundance behavior and carbon abundance behavior. This is
generally done by measuring the absorption strength of the 4300\,{\AA} CH
G-band and comparing with the CN bandstrengths. Previous studies of cluster
giants have shown that, at a given luminosity, the CN-strong stars tend to
have weak CH G-bands, implying nitrogen enhancement. Additionally,
\citet{gra00} demonstrated that the general behavior in clusters is that
the CH bandstrength decreases with increasing luminosity, while the summed
ratio [(C+N)/Fe] remains constant -- again indicating the presence of
nitrogen enhancement as stars move up the RGB. However, these abundances
are not predicted by standard stellar evolution models to change
significantly along the RGB between first dredge-up and helium flash
\citep{ibe64}, at least without the assumption of additional mixing.  

\begin{figure}
\centering
\plotone{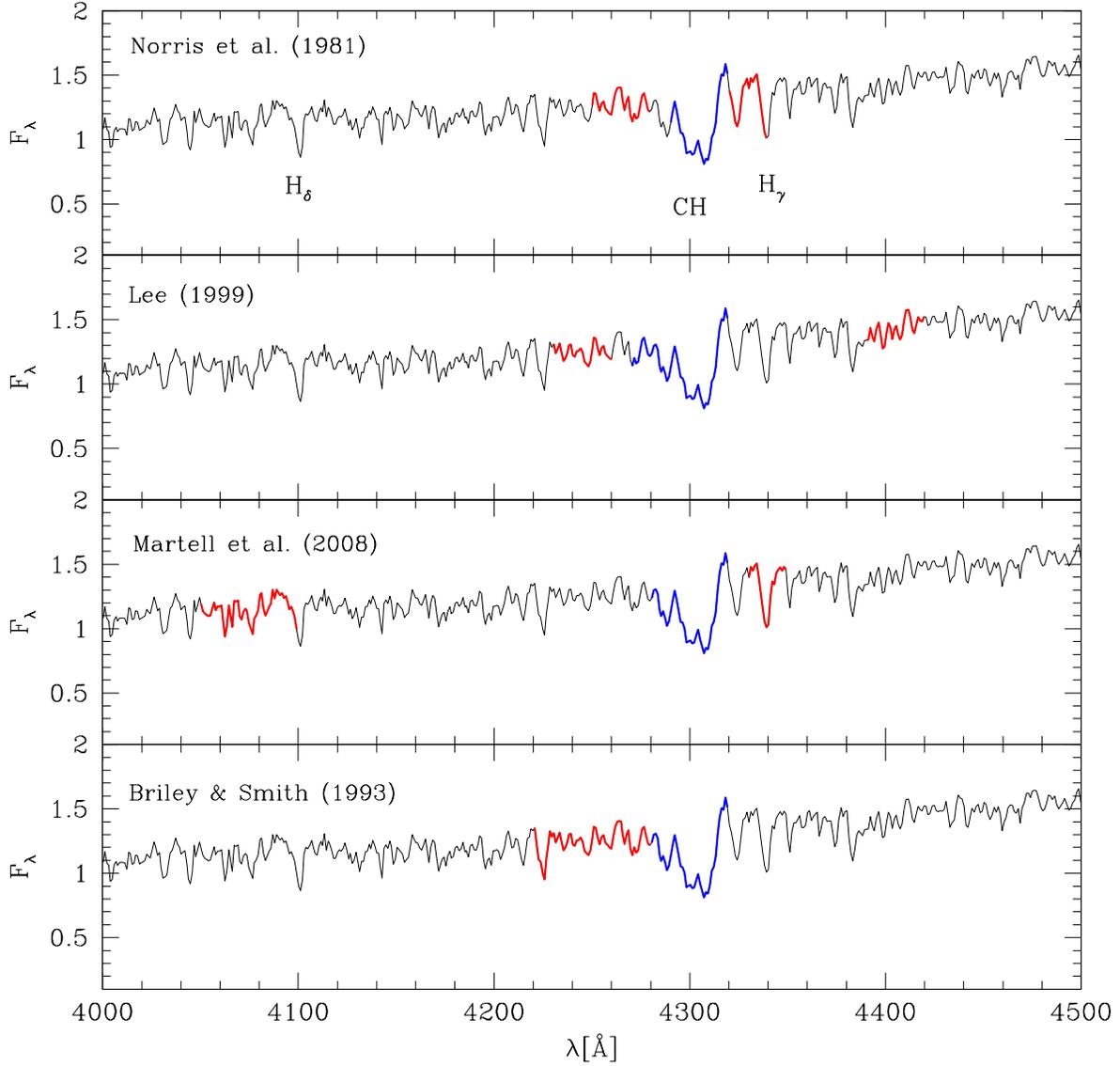}
\caption{CH bandstrength measurement windows for four popular index definitions, shown for a typical M3 red giant
spectrum (fiber 2475-53845-489).  The blue line on each spectrum indicates the line band while the red lines indicate the 
continuum windows used for comparison.  The CH G-band is indicated, along with two hydrogen Balmer lines for 
reference.}\label{figchspeccomp}
\end{figure}

Many definitions of indices that measure the CH G-band have been employed 
\citep{nor81a,tre83,bri93b,lee99,har03a,har03b,mar08a} and proposed
\citep{mar08b} in the past; these differ primarily because they were developed for
use with stars of specific metallicity or luminosity ranges. Figure
\ref{figchspeccomp} shows the linebands and sidebands used by
four common definitions. While the use of different CH bandstrength indices
complicates quantitative comparison with literature studies, qualitative
comparisons can still be useful. Inspection of Figure
\ref{figchspeccomp} reveals the presence of the H$_{\gamma}$ and H$_{\delta}$ Balmer lines
that can interfere with the continuum sidebands in the \citet{nor81a} and
\citet{mar08a} index definitions. These two definitions were developed
using samples of the most luminous, and therefore the coolest, red giants
in their clusters of interest, where Balmer absorption has minimal impact.
Of the two that remain, the \citet{bri93b} definition only includes the
blue side of the continuum, which can potentially cause artificially
depressed values for cooler stars where the continuum is more strongly
sloped. Therefore, the \citet{lee99} definition was adopted for this study:

\begin{equation}
CH(4300)_L = -2.5 \, \rm{log}\frac{\int_{4270}^{4320} I_{\lambda}\rm{d}\lambda}{\frac{1}{2}\left(\int_{4230}^{4260} I_{\lambda}\rm{d}\lambda + \int_{4390}^{4420} I_{\lambda}\rm{d}\lambda \right)}. \label{eqchlee99}
\end{equation}

\noindent This definition has the advantage that it has been used
for clusters covering a broad range of metallicities, from M15
\citep[$-$2.33;][]{lee00} to M71 \citep[$-$0.82;][]{lee05}, and it also
avoids the influence of Balmer lines that appear in hotter stars. In
addition, it samples the continuum on both sides of the CH G-band, thus
providing a more accurate representation of the ``expected'' continuum at
the location of the line band, regardless of the slope of the continuum in
this region.  The resulting indices are also tabulated in Table \ref{tabcnch}.

\begin{figure}
\plotone{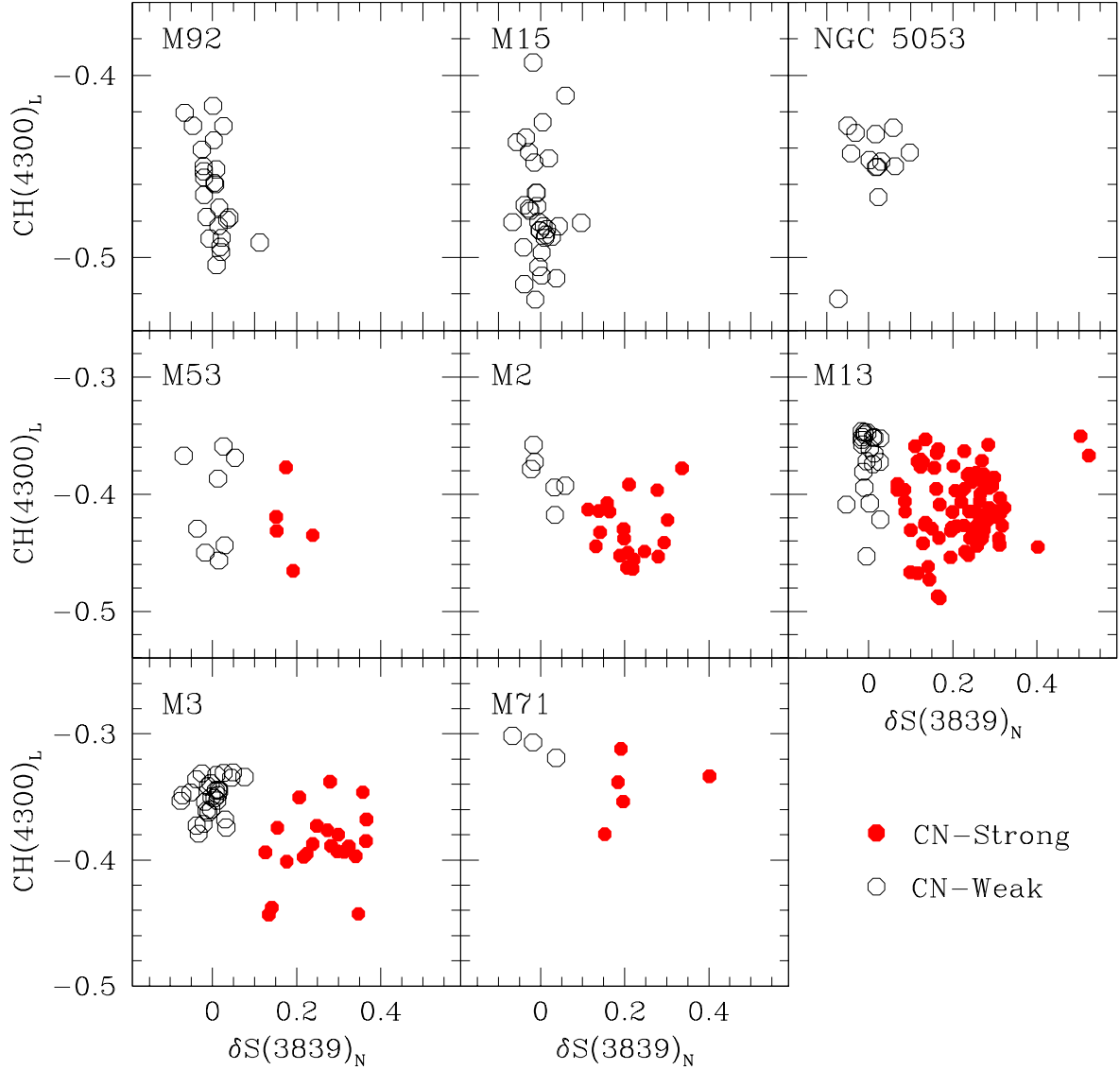}
\caption{CH absorption strength versus $\delta$S(3839)$_{\rm N}$ for RGB stars in our sample.  Red filled circles are 
CN-strong stars and black open circles are CN-weak stars. In most clusters, a trend from CH-strong/CN-weak to CH-weak/CN-strong is
seen.  The subscript `L' indicates the \citet{lee99} definition.}\label{figchdcn}
\end{figure}

\begin{figure}
\plotone{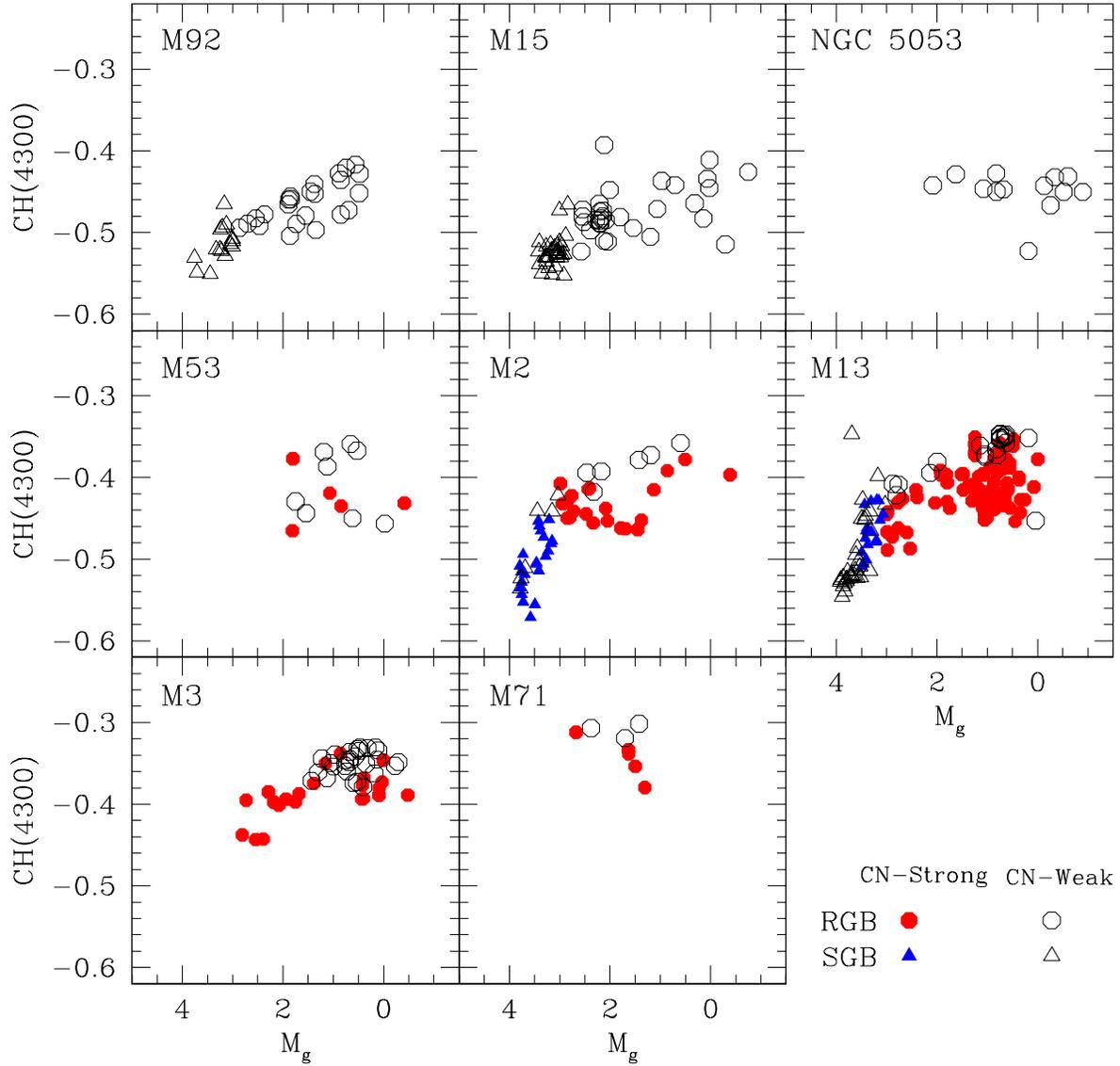}
\caption{CH absorption strength as a function of luminosity, with blue triangles corresponding to SGB stars while 
red circles are RGB stars.  Filled symbols represent CN-strong stars, while open symbols represent CN-weak 
stars.}\label{figchgcn}
\end{figure}

Figure \ref{figchdcn} shows the CH index versus $\delta$S(3839)$_{\rm N}$
for our cluster sample, ordered from low to high metallicity. From
inspection of this figure, CN-weak stars are typically also CH-strong,
while those stars that are CN-strong are typically CH-weak, with a few
exceptions. Figure \ref{figchgcn} is a similar set of plots, but here the
CH-index is plotted against absolute $g$-magnitude. The RGB stars within
each cluster exhibit the strongest CH absorption when they are CN-weak, and
vice versa. SGB stars are included as well for the two clusters in our
study where they are also available. Our results are consistent with
previous studies of giants in M2 \citep{smi90}, M13 \citep{sun81,bri93b,
smi96}, M3 \citep{sun81,smi96}, and M71 \citep{smi82a, lee05,alv08}.
Interestingly, M53 does not exhibit this anticorrelation in our sample or that
of \citet{mar08a}.

\subsection{CH Behavior at Low Metallicity}\label{subsecchlowfeh}

Examination of the CH-CN anticorrelation for low-metallicity clusters is
challenging, due in part to the decreased abundances of all metals in the
stellar atmospheres. This phenomenon has been demonstrated before with
synthetic spectra spanning a range of N abundances at low metallicity
\citep{mar08a}. This suggests a possible low-metallicity observational
cutoff, below which bimodality is either too difficult to detect with data
of the quality we were able to obtain or too difficult to measure due to
the low levels of CN present.

\begin{figure}
\plotone{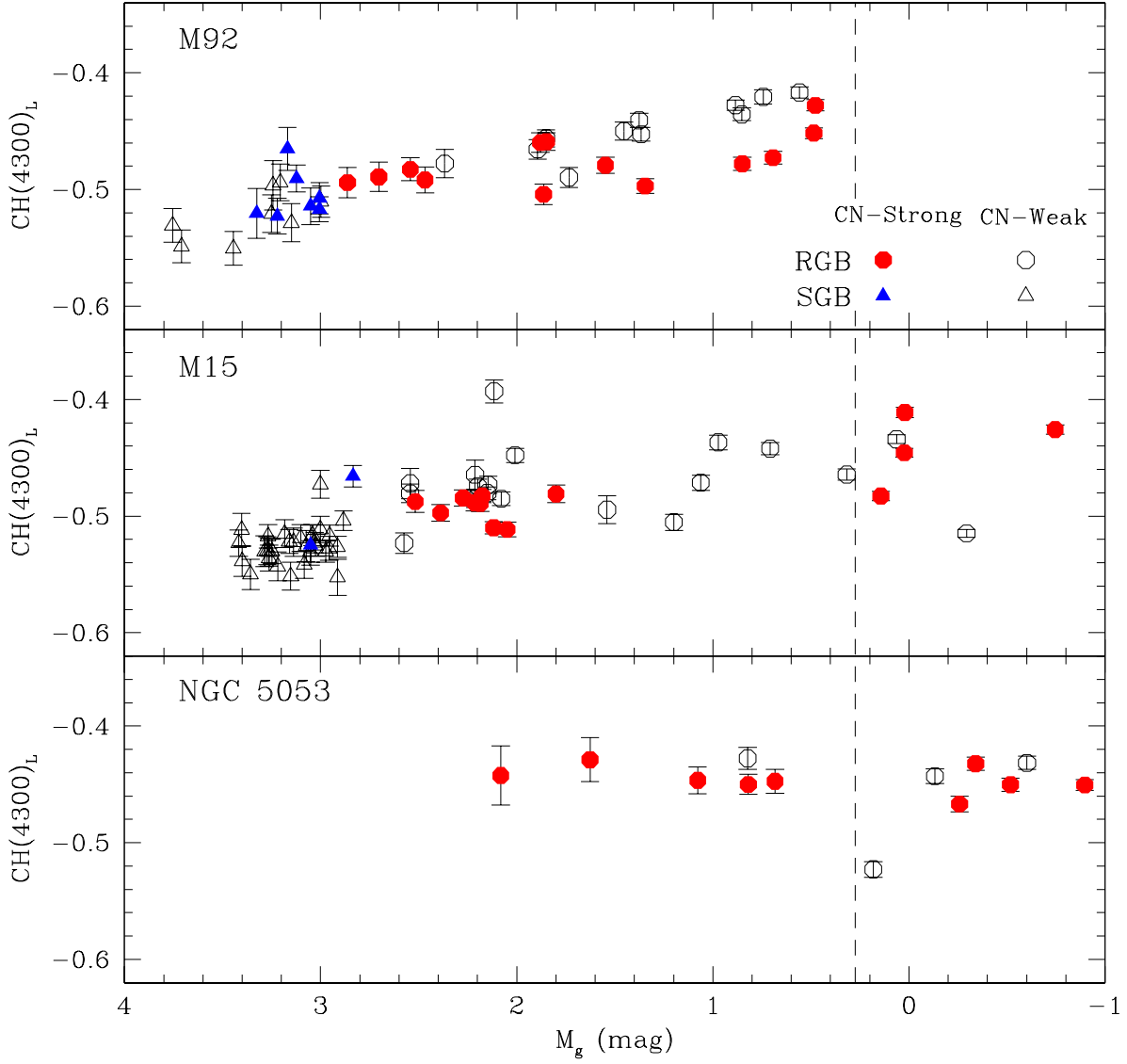}
\caption{Illustration of CH scatter on the RGBs of M92 (top), M15 (middle), and NGC~5053 (bottom).  
We have made a 
proposed CN-strong/weak cut at $\delta$S(3839)$_{\rm N}$=0.0, based on a
simple linear fit to the raw RGB S(3839)$_{\rm N}$ values, and have plotted
the CH abundance of CN-strong (filled circles) and CN-weak (open circles)
stars in each panel.  The vertical dashed line is the location of the RGB bump, drawn from
\citet{fus90}.}\label{figm92m15n5053chcndiv}
\end{figure}

\citet{she10} predict that abundance variations should exist at all metallicities,
even though bandstrength variations will become impossible to see at
sufficiently low values of \feh. To investigate this, we now consider
whether the CN-CH anticorrelation is observable in the very metal-poor
clusters M92, M15, and NGC~5053. Following the approach of
\citet{she10}, we fit a line to the raw S(3839)$_{\rm N}$ values as a function of
$g$-magnitude, then label those above the line as CN-strong and those below
the line as CN-weak. The results of this exercise for M92, M15, and
NGC~5053 are shown in Figure \ref{figm92m15n5053chcndiv}, plotted against
absolute $g$-magnitude for direct comparison. For the remainder of this
paper these two groups will be used to distinguish those stars with
slightly higher and slightly lower CN bandstrengths. At the lowest point on
the RGB (still above the SGB), there is little difference in the CH
bandstrengths between the CN-strong and CN-weak stars. However, as one
moves up the RGB, an apparent anticorrelation sets in for RGB stars in M92
at M$_g \approx 1.5$. Similar behavior is not seen in M15 and NGC~5053,
possibly due to the limited sampling.

\begin{figure}
\plotone{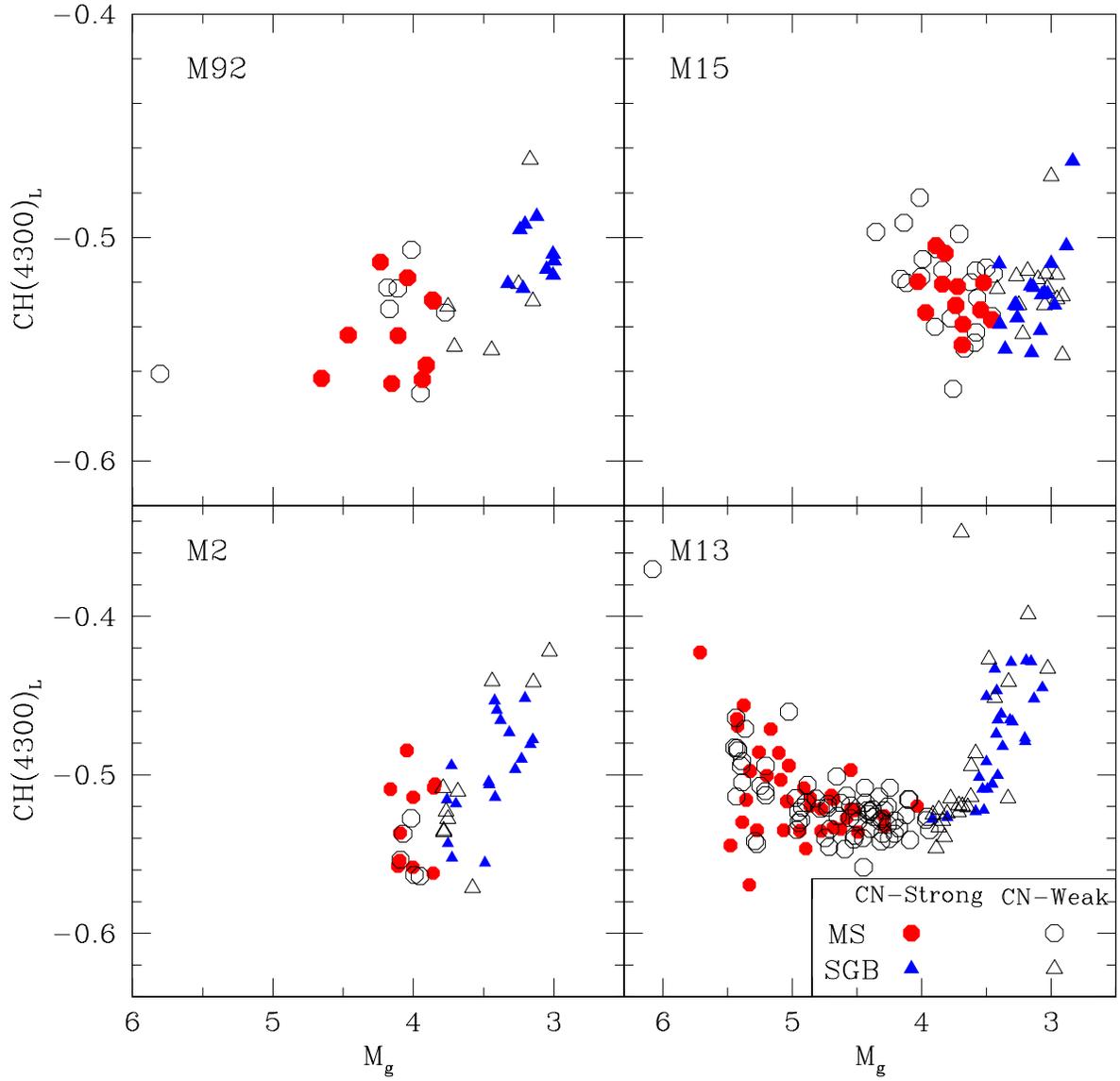}
\caption{CH absorption strength versus M$_g$ for MS and SGB stars.  Approximate divisions are 
made to indicate those stars that are higher (filled symbols) and lower
(open symbols) than the mean for CN absorption. No anticorrelation is
seen.}\label{figchgms}
\end{figure}

If CN abundance variations are in place prior to a star's evolution up the
RGB, one might expect to see an anticorrelation between CN and CH on the
SGB and MS. Although more difficult to observe, since higher temperatures
would tend to break up the molecules, such anticorrelations have been
reported previously for MS stars \citep{can98,coh99b,har03a,pan10}. Figure
\ref{figchgms} shows the relationship between CH and CN indices for MS and
SGB stars in M92, M15, M2, and M13, with the CN indices derived using the
\citet{har03a} definition. As in Fig. \ref{figdcnsnms}, we see a minimum in
CH bandstrength for M13 at M$_g \approx 4$, corresponding to the higher
effective temperatures of the turnoff point \citep[see Figure 2
from][]{bri04}. While an anticorrelation may exist on the upper SGBs of M2
and M13, which would be consistent with the signs of CN bimodality on the
SGB of these clusters and consistent with Figure \ref{figchgcn}, no other
anticorrelations are clearly seen. While further study is needed, we are
confident that follow-up observations of MS stars in these clusters will
reveal the same types of bandstrength variations as reported in previous
studies.

\section{Correlations with Cluster Parameters}\label{secclustcorr}

The possibility that CN enrichment is linked in some way with various physical
parameters of the parent cluster has been examined extensively in previous
studies. For example, \citet{nor87} identified a possible correlation
between CN bandstrength and the apparent ellipticity of the cluster using
data from 12 Galactic GCs. This suggested correlation was confirmed by
\citet{smi90}, who combined their observations of M2 with the set from
\citet{nor87}, but was not seen by \citet{kay08} using data from eight
clusters (of which three were in common between the two studies). They
concluded that the correlation with ellipticity is not as significant as
initially believed. \citet{smi90} also reported a correlation between CN
enrichment and cluster central velocity dispersion, which was again
disputed by \citet{kay08}. The \citet{smi90} claim of a CN enrichment
correlation with integrated cluster luminosity received moderate support
from the observations of \citet{kay08}.  

\begin{figure}
\centering
\plotone{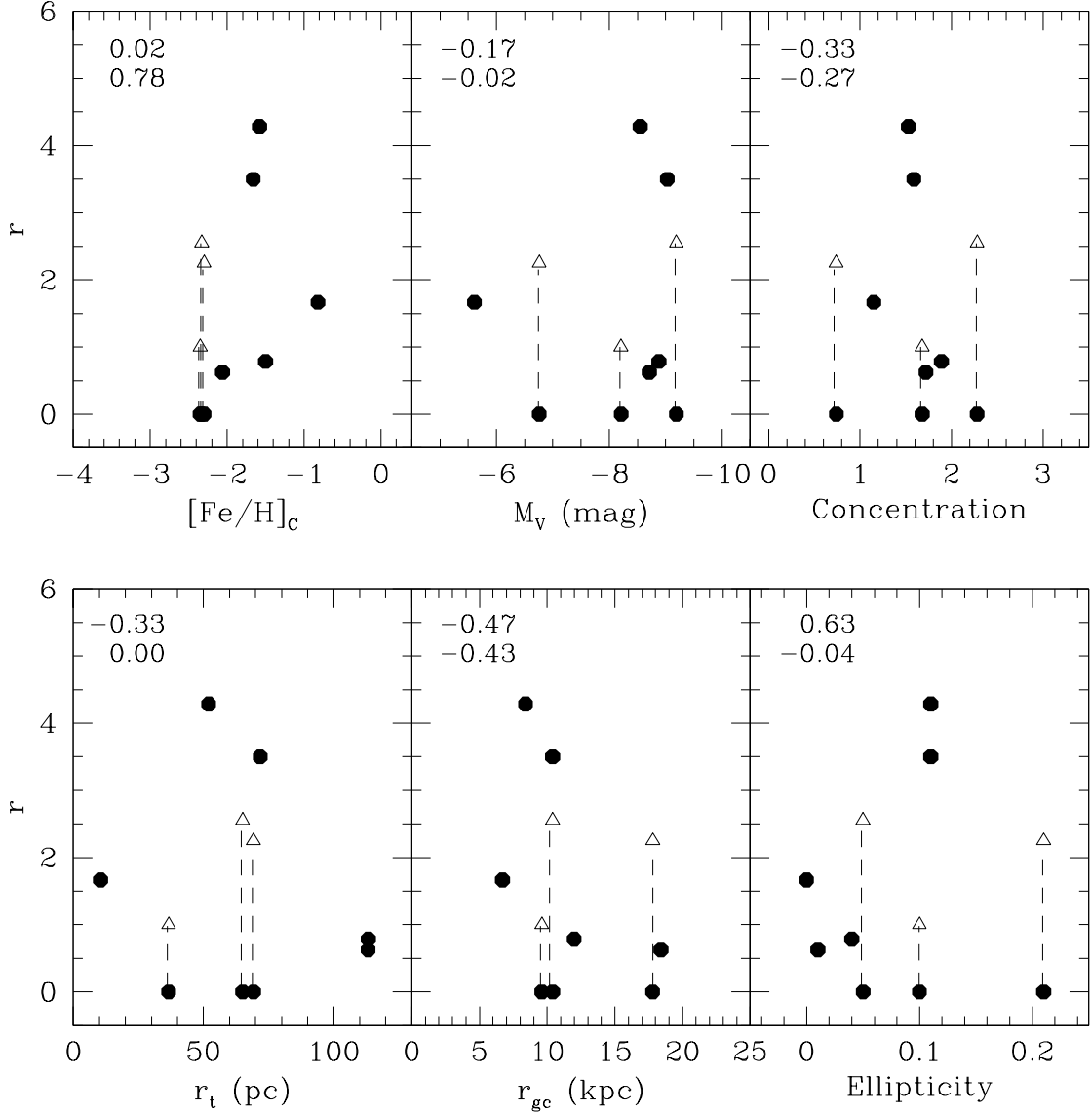}
\caption{The number ratio of CN-strong to CN-weak stars (designated as $r$) plotted against various cluster
parameters.  The filled circles represent the adopted ratios for each cluster, while the open triangles
representing M92 and M15 if we were to make the proposed divisions and identify relatively CN-strong and 
CN-weak stars in each.  The Spearman rank correlation coefficients are given in the upper-left corners of each
panel; the top numbers correspond to the correlation coefficient using the $r$ values from the proposed M92 and M15
divisions (triangles), while the bottom numbers correspond to the distribution using $r=0$ for M92 and M15.}\label{figcnratioparams}
\end{figure}

Figure \ref{figcnratioparams} shows the number ratio of CN-strong to
CN-weak RGB stars for the clusters in our sample, denoted by $r$, plotted
against various cluster parameters drawn from the \citet{har96}
catalog.\footnote{All references to \citet{har96} refer to the 2010 update
on his web page: {\tt http//www.physics.mcmaster.ca/$\sim$harris/mwgc.dat}.} These values are
tabulated in Table \ref{tabcnrationums}. The number ratio is useful because
it reveals the relative population sizes of the individual CN groups, and
provides a constraint on the chemical evolution of the cluster. 

M92 and M15 were treated differently because they do not possess CN-strong
stars, at least by the conventional definition, and hence would have an $r$
value of zero. However, Figures \ref{figm92cmddiv} and \ref{figm15cmddiv}
suggest the presence of stars that might be identifed as CN-strong
(relative to the rest of the cluster). Furthermore, an alternative method
for adopting relatively CN-strong and CN-weak stars was described in Section
\ref{subsecchlowfeh} for M92, M15, and NGC~5053. Using this approach, we determined
alternative $r$ values for M92, M15, and NGC~5053, which are plotted in Figure
\ref{figcnratioparams} as open triangles. Spearman rank correlation
coefficients were calculated for the distributions and are provided in the
upper-left corner of each panel; the top and bottom numbers correspond to
the sample with and without the second $r$ quantity included, respectively,
for M92, M15, and NGC~5053. 

\begin{figure}
\centering
\plotone{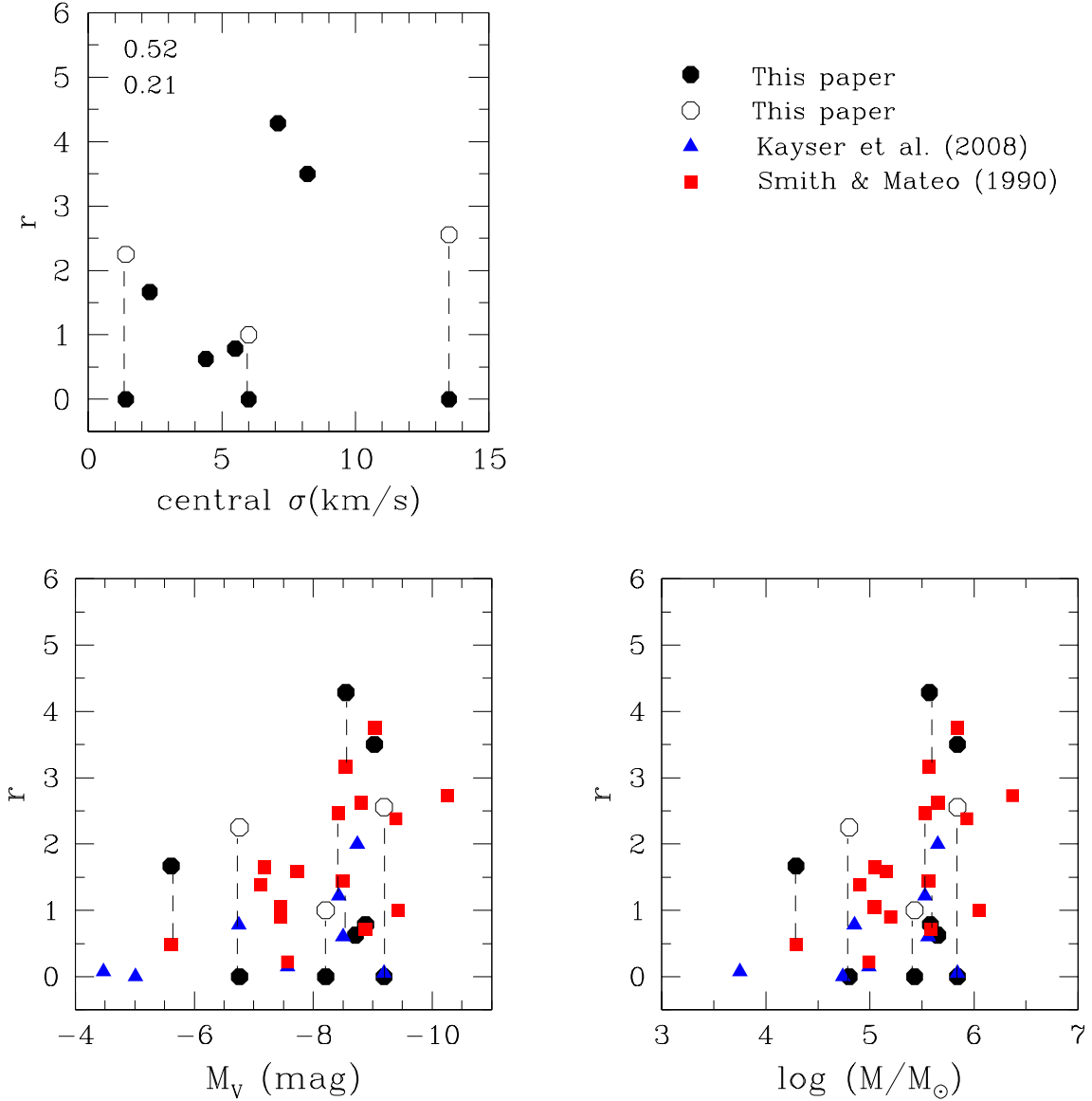}
\caption{The number ratio of CN-strong to CN-weak stars plotted against central velocity dispersion, absolute V magnitude, and
total cluster mass, combined with data from \citet{kay08} (blue triangles) and \citet{smi90} (red squares). 
The Spearman rank correlation coefficients for the $r$-$\sigma$ relation are given in the upper-left corners and correspond the same way as in Figure
\ref{figcnratioparams}.  A correlation is seen when using the CN division in M92 and M15.  Absolute magnitudes are drawn from the 2010
revision of the \citet{har96} catalog, while masses are drawn from \citet{man91} and \citet{mcl05}.  Data points that correspond to the same
clusters are connected with a dashed line.}\label{figcnratiosigMvMass}
\end{figure}

Supporting the claims of \citet{nor87} and \citet{smi90}, we observe a
moderate correlation between CN enrichment and cluster ellipticity. We
also note a moderate correlation between CN enrichment and central velocity dispersion
(Figure \ref{figcnratiosigMvMass}), in the same sense as \citet{smi90},
with a Spearman coefficient of 0.52 when using the ratios for the proposed
CN divisions in M92, M15, and NGC~5053, though no such correlation in our data is observed
with cluster luminosity (see Figure \ref{figcnratioparams}). However, the
range of cluster luminosities in our sample is not very broad. Only two
clusters with absolute $V$-magnitude brighter than $-$8.0 were observed,
one of which is metal-poor with no solidly identified CN-strong stars and the other
having fewer than 10 stars with reliable spectroscopy. When we combine our
data with that of \citet{kay08} and \citet{smi90}, also shown in Figure
\ref{figcnratiosigMvMass}, the overall trends become clearer; the largest
fraction of CN-strong stars are found in the most luminous and massive
clusters. This result is consistent with expectations from the
self-enrichment scenario -- the most massive clusters possess the deepest
gravitational potentials, allowing them to retain the largest amount of
chemically enriched gas expelled from evolving stars.

While studies to date typically quote an average value of the $r$ parameter
from the clusters in their samples, it is not clear that this quantity is
meaningful. Comparing the values directly is complicated by the fact that
some studies only use RGB stars, while others include subgiants, dwarfs,
and even AGB stars as well. Since dwarfs, with their hotter atmospheres,
are less likely to show significant CN absorption, their inclusion may bias
the value downward. The same holds true for the inclusion of AGB stars,
since they are nearly always CN-weak \citep[see][for further
discussion]{cam10}. The study by \citet{pan10}, comprised entirely of MS
dwarfs, reported an average of $r = 0.82 \pm 0.29$, while the average of the
$r$ values reported by \citet{kay08} for the giants in their sample is
0.61. Together, these results indicate that for the clusters in their
samples, both of which span a large range of \feh ~and luminosity, the
CN-strong stars are in the minority. However, studies of Na and O
abundances in cluster giants by \citet{car09b,car09a} suggest that the
ratio is much higher, with enriched stars comprising 50--70\% of the 
total ($r > 1$).
The compilation by \citet{smi90} of giants from 16 clusters gives an
average ratio of 1.72, which agrees well with \citet{car09b,car09a}. While
it is puzzling that two samples of cluster giants would yield such
discrepant results, this may result from a bias toward more luminous
clusters, since their inclusion would artificially inflate the fraction of
CN-strong stars in the sample. Table \ref{tabcnratioMv} lists the mean $r$
values alongside the mean $M_V$ values for our sample and the three
GCs from the literature. The increase in $\langle r \rangle$ with $\langle
M_V\rangle$ is apparent, indicating that conclusions drawn based upon the
CN ratio must account for any potential biases from including or excluding
massive GCs in the sample.

\section{Conclusions and Discussion}\label{secconcl}

We have used low-resolution SEGUE spectra to confirm the presence of a
bimodality in the CN distributions for stars in GCs with \feh ~$> -2$. In
addition, we extend the metallicity limit for which CN bimodality has been
observed to at least \feh ~$\sim -2.1$, by adding M53 to the collection.
Furthermore, we have presented evidence suggesting the presence of a much
smaller division between CN-strong and CN-weak groups on the RGB of M92,
down to luminosities corresponding to $M_V$ ~$\sim 1.8$, which is in
quantitative agreement with earlier studies suggesting carbon depletion
setting in below the RGB bump. Evidence for two CN groups on the RGB of
M15, with a similar metallicity as M92, also exists. Previous CN abundance
studies of M92 and M15 have not shown strong evidence for bimodality in
either cluster. We also confirm an overall anticorrelation between CN and
CH bandstrengths on the RGB for M2, M13, and M3, for luminosities beyond
the point of first dredge-up, while offering evidence that M92 may also
display the same anticorrelation. Our samples for M53 and M71 are too small
to make strong claims for anticorrelations, although M53 appears to have CN
and CH bandstrengths that are uncorrelated. Despite its chemical similarity
to M92, no apparent CN-CH anticorrelation is present among M15 giants.
Finally, NGC~5053 exhibits a remarkably uniform CH bandstrength along the
RGB, in spite of the fact that our sample of stars straddles the RGB bump,
where deep mixing is predicted to set in.

Our observations support a scenario in which evolved stars disperse
enriched gas throughout the cluster that ultimately forms a second
generation of stars. This results in two chemically disparate populations
of stars. Such abundance variations are not observed among even the oldest
and most massive open clusters \citep{jac08,mar09}, presumably due to their
significantly lower gravitational potentials preventing them from retaining
enriched gas from evolved stars. Furthermore, current theoretical models of
the origin of light-element abundance variations rely heavily on the high
density environments of proto-globular clusters to facilitate enrichment of
subsequent stellar populations. Since these chemically enriched stellar
populations appear to form preferentially in GCs, this motivates the
argument that the presence of CN-strong stars in the Galactic halo may have
been stripped from GCs into the field, rather than being the result of
\emph{in situ} formation \citep{mar10}. Additional studies of halo GCs
should provide an opportunity to explore and quantify this contribution
directly.

\acknowledgements

J.P.S., T.C.B., and Y.S.L. acknowledge partial funding of this work from grants
PHY 02-16783 and PHY 08-22648: Physics Frontiers Center/Joint Institute
for Nuclear Astrophysics (JINA), awarded by the US National Science
Foundation.

% Any other funding acknowledgements?  DCF?

Funding for the SDSS and SDSS-II has been provided by the Alfred P.
Sloan Foundation, the Participating Institutions, the National
Science Foundation, the U.S. Department of Energy, the National
Aeronautics and Space Administration, the Japanese Monbukagakusho,
the Max Planck Society, and the Higher Education Funding Council for
England. The SDSS Web Site is {\tt http://www.sdss.org/}.

The SDSS is managed by the Astrophysical Research Consortium for the
Participating Institutions. The Participating Institutions are the American
Museum of Natural History, Astrophysical Institute Potsdam, University of
Basel, University of Cambridge, Case Western Reserve University, University
of Chicago, Drexel University, Fermilab, the Institute for Advanced Study,
the Japan Participation Group, Johns Hopkins University, the Joint
Institute for Nuclear Astrophysics, the Kavli Institute for Particle
Astrophysics and Cosmology, the Korean Scientist Group, the Chinese Academy
of Sciences (LAMOST), Los Alamos National Laboratory, the
Max-Planck-Institute for Astronomy (MPIA), the Max-Planck-Institute for
Astrophysics (MPA), New Mexico State University, Ohio State University,
University of Pittsburgh, University of Portsmouth, Princeton University,
the United States Naval Observatory, and the University of Washington.

\clearpage
\begin{deluxetable}{llcccccc}
\tabletypesize{\scriptsize}
\tablecolumns{8}
\tablewidth{0pt}
\tablecaption{Globular Cluster Photometric Properties}
\tablehead{\colhead{Cluster} & & \colhead{RA (J2000)} & \colhead{Dec (J2000)} & \colhead{$(l, b)$} & \colhead{$(m-M)_{0}^{\rm{a}}$} & \colhead{$E(B-V)^{\rm{b}}$} & \colhead{\rm{r}$_{t}^{\rm{a}}$} \\
 & & & & & & & \colhead{\rm{(arcmin)}}}
\startdata
NGC 6341 & M92 & 17:17:07.39 & $+$43:08:09.4 & (68.3, $+$34.9)  & 14.64 & 0.023 & 15.17 \\
NGC 7078 & M15 & 21:29:58.33 & $+$12:10:01.2 & (65.0, $-$27.3)  & 15.37 & 0.110 & 21.50 \\
NGC 5053 &     & 13:16:27.09 & $+$17:42:00.9 & (335.7, $+$78.9) & 16.12 & 0.017 & 13.67 \\
NGC 5024 & M53 & 13:12:55.25 & $+$18:10:05.4 & (333.0, $+$79.8) & 16.25 & 0.021 & 21.75 \\
NGC 7089 & M2  & 21:33:27.02 & $-$00:49:23.7 & (58.4, $-$35.8)  & 15.49 & 0.045 & 21.45 \\
NGC 6205 & M13 & 16:41:41.24 & $+$36:27:35.5 & (59.0, $+$40.9)  & 14.48 & 0.017 & 25.18 \\
NGC 5272 & M3  & 13:42:11.62 & $+$28:22:38.2 & (42.2, $+$78.7)  & 14.95 & 0.013 & 38.19 \\
NGC 6838 & M71 & 19:53:46.49 & $+$18:46:45.1 & (56.7, $-$4.6)   & 12.86 & 0.275$^{\rm{c}}$ & 8.96 \\
\enddata
\tablecomments{Photometric properties of the globular clusters in our sample as drawn from the literature.  The 
parameter r$_{t}$ is the tidal radius in arcminutes.  The listed distance modulus $(m-M)_0$ is extinction corrected.
Note that the Harris values have been updated from {\tt http//www.physics.mcmaster.ca/$\sim$harris/mwgc.dat}. \\
References: $^{\rm{a}}$ \citet{har96}; $^{\rm{b}}$ \citet{sch98}; $^{\rm{c}}$ \citet{gru02b}.}
\label{tabclusterphotprops}
\end{deluxetable}

\clearpage
\begin{deluxetable}{lcccccccc}
\tabletypesize{\scriptsize}
\tablecolumns{9}
\tablewidth{0pt}
\tablecaption{Globular Cluster Spectroscopic and Physical Properties}
\tablehead{\colhead{Cluster} & \colhead{$\rm{[Fe/H]^{a}}$} & \colhead{$\rm{[Fe/H]_C}$} & \colhead{\rm{V}$_{r}^{\rm{a}}$} & \colhead{log($\frac{M}{M_{\odot}}$)} & \colhead{R$^{\rm{a}}_{\rm{gc}}$} & \colhead{$C^{\rm{a}}$} & \colhead{$\sigma^{\rm{a}}$} & \colhead{$\varepsilon^{\rm{a}}$} \\
 & & & \colhead{$\rm{(km~s^{-1})}$} & \colhead{(dex)} & \colhead{(kpc)} & & \colhead{$\rm{(km~s^{-1})}$} & }
\startdata
M92      & $-$2.31 & $-$2.35 & $-$120.0 & $5.43^{b}$ & 9.6  & 1.68 & 6.0  & 0.10 \\
M15      & $-$2.37 & $-$2.33 & $-$107.0 & $5.84^{c}$ & 10.4 & 2.28 & 13.5 & 0.05 \\
NGC 5053 & $-$2.27 & $-$2.30 & $+$44.0  & $4.80^{b}$ & 17.8 & 0.74 & 1.4  & 0.21 \\
M53      & $-$2.10 & $-$2.06 & $-$62.9  & $5.65^{b}$ & 18.4 & 1.72 & 4.4  & 0.01 \\
M2       & $-$1.65 & $-$1.66 & $-$5.3   & $5.84^{b}$ & 10.4 & 1.59 & 8.2  & 0.11 \\
M13      & $-$1.53 & $-$1.58 & $-$244.2 & $5.57^{b}$ & 8.4  & 1.53 & 7.1  & 0.11 \\
M3       & $-$1.50 & $-$1.50 & $-$147.6 & $5.58^{b}$ & 12.0 & 1.89 & 5.5  & 0.04 \\
M71      & $-$0.78 & $-$0.82 & $-$22.8  & $4.29^{c}$ & 6.7  & 1.15 & 2.3  & 0.00 \\
\enddata
\tablecomments{Spectroscopic and physical properties of the globular clusters in our sample as drawn from the literature.  The 
parameter $\rm{[Fe/H]_C}$ is from the recalibrated globular cluster metallicity scale of \citet{car09}. 
Distance from Galactic center R$_{\rm{gc}}$ calculated assuming R$_0$ = 8.0 kpc. Central concentration $C$ 
derived from a King model where $C$ = log(r$_t$/r$_c$). 
$^{\rm{c}}$ \citet{man91}. Note that the Harris values have been updated from {\tt http//www.physics.mcmaster.ca/$\sim$harris/mwgc.dat}. \\
References: $^{\rm{a}}$ \citet{har96}; $^{\rm{b}}$ \citet{mcl05};}
\label{tabclusterspecprops}
\end{deluxetable}

\clearpage
%\begin{landscape}
\begin{deluxetable}{lcccrcrrcc}
\tabletypesize{\scriptsize}
\renewcommand{\tabcolsep}{2pt}
\tablecolumns{7}
\tablewidth{0in}
\tablecaption{Line Indices of Adopted True Member Stars}
\tablehead{
\colhead{spSpec name} & \colhead{\gr} & \colhead{$g$} & \colhead{$\sigma_{g}$} &
\colhead{S(3839)} & \colhead{$\sigma_{\rm{S(3839)}}$} & \colhead{$\delta$S(3839)} & 
\colhead{CH(4300)$_{\rm{L}}$} & \colhead{$\sigma_{\rm{CH(4300)}}$} & \colhead{Type} \\

\colhead{(1)} & \colhead{(2)} & \colhead{(3)} & \colhead{(4)} & \colhead{(5)} & \colhead{(6)} & \colhead{(7)} &
\colhead{(8)} & \colhead{(9)} & \colhead{(10)}
}

\startdata
 & & & M92 & & & \\
2247-54169-362 & 0.441 & 17.343 & 0.011 & 	$-$0.172 & 0.031 & 	 0.022 & 	$-$0.489 & 0.013 & RGB \\ 
2247-54169-364 & 0.456 & 16.533 & 0.015 & 	$-$0.196 & 0.015 & 	$-$0.020 & 	$-$0.466 & 0.008 & RGB \\ 
2247-54169-367 & 0.439 & 16.519 & 0.009 & 	$-$0.169 & 0.018 & 	 0.006 & 	$-$0.460 & 0.008 & RGB \\ 
2247-54169-380 & 0.484 & 16.093 & 0.010 & 	$-$0.187 & 0.016 & 	$-$0.021 & 	$-$0.450 & 0.008 & RGB \\ 
2247-54169-404 & 0.455 & 16.503 & 0.010 & 	$-$0.165 & 0.016 & 	 0.010 & 	$-$0.504 & 0.009 & RGB \\ 
2247-54169-444 & 0.467 & 16.006 & 0.011 & 	$-$0.184 & 0.010 & 	$-$0.020 & 	$-$0.453 & 0.006 & RGB \\ 
2247-54169-451 & 0.472 & 16.188 & 0.008 & 	$-$0.133 & 0.013 & 	 0.035 & 	$-$0.479 & 0.007 & RGB \\
\enddata
\tablecomments{
{L}ine index values derived from adopted cluster member stars. Column 1 lists the spSpec name, which identifies the star
on the spectral plate in the form of spectroscopic plug-plate number (four digits), Modified Julian Date (five digits),
and fiber used (three digits), while Columns 2, 3, and 4 list the \gr ~and $g_0$ photometry.  Column 5 lists the raw 
S(3839) CN index, which for AGB, RGB, and SGB stars was calculated
using the definition from \citet{nor81a} and for MS stars was calculated using the definition from \citet{har03a}, and
column 6 lists the corresponding uncertainties.  Column 7 gives the temperature-corrected CN index, and columns 8 and 9
list the CH indices and corresponding uncertainties. Column 10 indicates the luminosity class. \\
(This table is available in its entirety in machine-readable and Virtual Observatory (VO) forms
in the online journal.  A portion is shown here for guidance regarding its form and content.)
}\label{tabcnch}
\end{deluxetable}
%\end{landscape}

\clearpage
\begin{deluxetable}{lccc}
\tabletypesize{\scriptsize}
\tablecolumns{4}
\tablewidth{0pt}
\tablecaption{Individual Cluster RGB CN Number Ratios}
\tablehead{\colhead{Cluster} & \colhead{N$_{s}$} & \colhead{N$_{w}$} & \colhead{$r$} }
\startdata
M92      & 0 (12) & 24 (12) & 0.00 (1.00) \\
M15      & 0 (23) & 32 (9)  & 0.00 (2.56) \\
NGC~5053 & 0 (9)  & 13 (4)  & 0.00 (2.25) \\
M53      & 5      & 8       & 0.62 \\
M2       & 21     & 6       & 3.50 \\
M13      & 90     & 21      & 4.29 \\
M3       & 22     & 28      & 0.79 \\
M71      & 5      & 3       & 1.67 \\
\enddata
\tablecomments{Number of CN-strong (N$_{s}$) and CN-weak (N$_{w}$) stars, and CN number ratios, for RGB stars 
from individual GCs in our sample. Values in parentheses for M92 and M15
represent the groupings based on the proposed divisions described in the
text.}
\label{tabcnrationums}
\end{deluxetable}

\clearpage
\begin{deluxetable}{lccc}
\tabletypesize{\scriptsize}
\tablecolumns{4}
\tablewidth{0pt}
\tablecaption{Literature CN Number Ratios}
\tablehead{\colhead{Sample} & \colhead{N} & \colhead{$\langle M_V \rangle$} & \colhead{$\langle r \rangle$}}
\startdata
\citet{kay08} & 8  & $-6.19$ & 0.61 \\
\citet{pan10} & 12 & $-7.84$ & 0.82 \\
This study    & 8  & $-8.12$ & 1.36 \\
\citet{smi90} & 16 & $-8.21$ & 1.72 \\
\enddata
\tablecomments{Mean CN number ratios from four samples of GCs as a function of absolute integrated magnitude.
$N$ indicates the number of clusters included in each sample.}\label{tabcnratioMv}
\end{deluxetable}

\clearpage

\clearpage

\clearpage

\clearpage

\clearpage

\clearpage

\clearpage

\clearpage

\clearpage

\clearpage

\clearpage

\clearpage

\clearpage

\clearpage

\clearpage

\clearpage

\clearpage

\clearpage

\clearpage

\clearpage

\clearpage

\clearpage

\end{document}